\documentclass[aps,prd,preprint,nofootinbib,11pt]{revtex4}
\usepackage{amsfonts}
\usepackage{mathrsfs}
\usepackage{graphicx}
\usepackage{amsmath}
\usepackage{amssymb}
\usepackage{subfigure}
\usepackage{epsfig}
\usepackage{graphicx}
\usepackage{color}
\newcommand{\bqa}{\begin{eqnarray}}
\newcommand{\eqa}{\end{eqnarray}}
\newcommand{\beq}{\begin{equation}}
\newcommand{\eeq}{\end{equation}}
\allowdisplaybreaks[1]
\graphicspath{{fig/}{dia/}} \DeclareGraphicsExtensions{.eps}

\begin{document}
\title{Light baryonium spectrum}

\author{Bing-Dong Wan$^{1,2}$\footnote{wanbingdong16@mails.ucas.ac.cn}, Sheng-Qi Zhang$^1$\footnote{zhangshengqi20@mails.ucas.ac.cn} and Cong-Feng Qiao$^{1,3}$\footnote{qiaocf@ucas.ac.cn, corresponding author}\vspace{+3pt}}

\affiliation{$^1$ School of Physics, University of Chinese Academy of Science, Yuquan Road 19A, Beijing 100049, China \\
$^2$ School of Fundamental Physics and Mathematical Sciences, Hangzhou Institute for Advanced Study, UCAS, Hangzhu 310024, China\\
$^3$ CAS Center for Excellence in Particle Physics, Beijing 100049, China}

\author{~\\~\\}

\begin{abstract}
\vspace{0.3cm}
We evaluate the light baryonium spectrum, viz. the baryon-antibaryon states, in the framework of QCD sum rules. The nonperturbative contributions up to dimension 12 are taken into account. Numerical results indicate that there might exist eight possible light baryonium states, i.e. $p$-$\bar{p}$, $\Lambda$-$\bar{\Lambda}$, $\Sigma$-$\bar{\Sigma}$, and $\Xi$-$\bar{\Xi}$ with quantum numbers of $0^{-+}$  and $1^{--}$. For the $\Lambda$-$\bar{\Lambda}$, $\Sigma$-$\bar{\Sigma}$, and $\Xi$-$\bar{\Xi}$ states, their masses are found above the corresponding dibaryon thresholds, while the masses of $p$-$\bar{p}$ states are not. The possible baryonium decay modes are analyzed, which are hopefully measurable in BESIII, BELLEII, and LHCb experiments.
\end{abstract}
 \maketitle
\newpage

\section{Introduction}

The establishment of quark model (QM) in 1960s \cite{GellMann:1964nj,Zweig} led to a renaissance in the exploration of micro worlds. The spectroscopy of conventional hadrons ($q\bar{q}$ or $qqq$) in QM is being gradually confirmed through experiments and will be completed soon. Entering the new millennium, with development of technology, the emergence of the so-called exotic state such as X(3872) has been reported \cite{Choi:2003ue}, and new ones tend to appear more frequently. Presently, a bunch of charmoniumlike/bottomoniumlike states XYZ and pentaquark states $P_c$ are observed in experiments; this situation is similar to the phase of “particle zoo” witnessed in the last century. To discover more exotic states and explore their properties is currently one of the most intriguing and important topics in particle physics, which may greatly enrich the hadron family and our knowledge of the nature of QCD.

 Facing the observations of tetraquark and pentaquark states, it is nature to conjecture existence of the hexaquark states, and it is time for hunting them. Deuteron, created at the beginning of the Universe and its stability is responsible for the production of other elements, is a typical and well-established dibaryon molecular state with $J^P=1^+$ and binding energy $E_B = 2.225 \text{MeV}$ \cite{Weinberg:1962hj}. Interestingly, the strong interactions bring stability to deuterons and also allow various other stable deuteronlike dibaryon states;  however, no such states, though speculated about many times \cite{Jaffe:1976yi,Mulders:1980vx,Balachandran:1983dj,Sakai:1999qm,Ikeda:2007nz,Bashkanov:2013cla,Shanahan:2011su,Clement:2016vnl}, have been observed yet \cite{BaBar:2018hpv}. On the other hand, the baryonium state composed of a baryon-antibaryon pair is another special class of heaxquark configuration. The interaction between baryon-antibaryon pair  is analogous to that between two baryons, which will provide important hints to understanding the absence in observation of the stable deuteronlike dibaryon states.

Actually, the history of the investigation of baryon-antibaryon states dates back to the 1940s, when Fermi and Yang proposed that $\pi$-mesons may be composite particles formed by the association of a nucleon with an antinucleon \cite{Fermi:1949voc}, and their scenario was later on replaced by the quark model. Entering the new millennium, the heavy baryon-antibaryon hadronic structures, and, hence, the term of baryonium, were proposed and employed to explain the extraordinary nature of $Y(4260)$ \cite{Qiao:2005av, Qiao:2007ce} and other charmoniumlike states observed in experiments. Later on, more investigations on baryonium are performed from various aspects \cite{Chen:2011cta, Chen:2013sba,Wan:2019ake,Chen:2016ymy,Liu:2007tj,Wang:2021qmn}. Partly due to the the small spacings between light hadron states, it is normally hard to discriminate exotic states from conventional ones. However, with a large amount of $J/\psi$ samples, BESIII Collaboration is carefully examining the physics happening in the energy region around $2.0$ GeV \cite{BESIII:2010gmv,BES:2003aic,BES:2005ega,BESIII:2010vwa,BESIII:2019wkp,BESIII:2016qzq,BESIII:2020vtu, BESIII:2017kqw,BESIII:2017hyw,BESIII:2019cuv}, which motivates a fresh interest in light exotic states . In the literature, theoretical investigations on light baryonium were made through various techniques, including flux tube model \cite{Deng:2012wi,Deng:2013aca}, one-boson-exchange potential (OBEP) model \cite{Zhao:2013ffn}, Bethe-Salpeter approach \cite{Zhu:2019ibc,Wang:2010vz}, and QCD sum rules (QCDSR) \cite{Wang:2006sna}.

Of those techniques, the model with the independent Shifman, Vainshtein, and Zakharov (SVZ) sum rule technique \cite{Shifman} has some peculiar advantages in exploring hadron properties involving nonperturbative QCD. Rather than a phenomenological model, QCDSR is a QCD based theoretical framework which incorporates nonperturbative effects universally order by order and has already achieved a lot in the study of hadron spectroscopy and decays. Its starting point is to construct the proper interpolating currents corresponding to the hadron of interest ,which possesses the foremost information about the concerned hadron, like quantum numbers and the constituent quark or gluon. With the currents, the two-point correlation function, which has two representations, the QCD representation and the phenomenological representation, can be constructed. Equating these two representations, the QCD sum rules will be formally established, from which the hadron mass and decay width may be deduced.

In this work, we investigate the light baryonium states, i.e., $\mathscr{B}$-$\bar{\mathscr{B}}$ states with quantum numbers $J^{PC }= 0^{++}$, $0^{-+}$, $1^{++}$ and $1^{--}$ in the framework of QCD sum rules, where $\mathscr{B}$ denotes light baryon. The possible baryonium decay channels are also analyzed. The rest of the paper is organized as follows. After the introduction, a brief interpretation of QCD sum rules and some primary formulas in our calculation are presented in Sec. \ref{Formalism}. We give the numerical analysis and results in Sec. \ref{Numerical}. In Sec. \ref{Decay}, possible decay modes of light baryonium states are investigated. The last part is left for conclusions and discussions.

\section{Formalism}\label{Formalism}

To evaluate the mass spectrum of $\mathscr{B}$-$\bar{\mathscr{B}}$ states in QCDSR, the appropriate currents coupling to the states have to be constructed. The lowest order interpolating currents for $\mathscr{B}$-$\bar{\mathscr{B}}$ states with quantum numbers $ 0^{++}$, $0^{-+}$,  $1^{++}$, and $1^{--}$ can be respectively constructed as
\begin{eqnarray}\label{current_lambda}
j^{0^{-+}}(x)&=& i\,\bar{\eta}_{\mathscr{B}}(x) \gamma_5 \eta_{\mathscr{B}}(x) \;,\label{Ja0-+}\\
j^{1^{--}}_\mu(x)&=&\bar{\eta}_{\mathscr{B}}(x) \gamma_\mu \eta_{\mathscr{B}}(x)\;, \label{Ja1--}\\
j^{0^{++}}(x)&=& \bar{\eta}_{\mathscr{B}}(x) \eta_{\mathscr{B}}(x)\;,\label{Ja0++}\\
j^{1^{++}}_\mu(x)&=& i\,\bar{\eta}_{\mathscr{B}}(x) \gamma_\mu\gamma_5 \eta_{\mathscr{B}}(x) \;. \label{Ja1++}
\end{eqnarray}
Here, we use the notion $\eta_{\mathscr{B}}$ to represent the Dirac baryon fields  without free Lorentz indices. As shown in Ref. \cite{Chung:1981wm}, $\eta_{\mathscr{B}}$ may takes the following quark structure :
\begin{eqnarray}\label{current_lambda_b}
\eta_{\mathscr{B}}(x)&=&i \epsilon_{a b c}[ q_a^{i\,T}(x) C \gamma_5 q_b^j(x) ]q^k_c(x) \; ,
\end{eqnarray}
where the superscripts $i$, $j$, and $k$ denote the flavor of light quarks, subscripts $a$, $b$, and $c$ are color indices, and $C$ is the charge conjugation matrix. In our calculation, $(i, j, k)=(u, d, u)$, $(u, d, s)$, $(u, s, d)$, and (s, u, s) for $p$, $\Lambda$, $\Sigma$ and $\Xi$ states, respectively.

With the currents (\ref{Ja0-+})$-$(\ref{Ja1++}), the two-point correlation function can be readily established, i.e.,
\begin{eqnarray}
\Pi(q^2) &=& i \int d^4 x e^{i q \cdot x} \langle 0 | T \{ j (x),\;  j^\dagger (0) \} |0 \rangle\;,\\
\Pi_{\mu\nu}(q^2) &=& i \int d^4 x e^{i q \cdot x} \langle 0 | T \{ j_\mu (x),\;  j_\nu^\dagger (0) \} |0 \rangle \;
\end{eqnarray}
where $j(x)$ and $j_\mu(x)$ are the relevant hadronic currents with $J = 0$ and 1, respectively, and $|0\rangle$ denotes the physical vacuum. For $j_\mu(x)$, the correlation function has the following Lorentz covariance form:
\begin{eqnarray}
\Pi_{\mu\nu}(q^2) &=&-\Big( g_{\mu \nu} - \frac{q_\mu q_\nu}{q^2}\Big) \Pi_1(q^2)+ \frac{q_\mu q_\nu}{q^2}\Pi_0(q^2)\;,
\end{eqnarray}
where the subscripts $1$ and $0$, respectively, denote the quantum numbers of the spin 1 and 0 mesons.

On the phenomenological side, after separating the ground state contribution from the hadronic state, the correlation function $\Pi^{X}_{J^{PC}}(q^2)$ can be expressed as a dispersion integral over the physical regime, i.e.,
 \begin{eqnarray}
\Pi^{X,\;phen}_{J^{PC}}(q^2) & = & \frac{(\lambda^X_{J^{PC}})^2}{(m^X_{J^{PC}})^2 - q^2} + \frac{1}{\pi} \int_{s_0}^\infty d s \frac{\rho^X_{J^{PC}}(s)}{s - q^2} \; , \label{hadron}
\end{eqnarray}
where the superscript $X$ denotes the lowest lying  $\mathscr{B}$-$\bar{\mathscr{B}}$ hexaquark state, $m^X_{J^{PC}}$ is the mass of $X$ with the quantum number of $J^{PC}$, $\rho^X_{J^{PC}}(s)$ is the spectral density that contains the contributions from higher excited states and the continuum states above the threshold $s_0$, and $\lambda^X$ is the coupling constant.

On the OPE (operator product expansion) side, the dispersion relation can express the correlation function $\Pi^{X}_{J^{PC}}(q^2)$ as
 \begin{eqnarray}
\Pi^{X,\;OPE}_{J^{PC}} (q^2) &=& \int_{s_{min}}^{\infty} d s
\frac{\rho^{X,\;OPE}_{J^{PC}}(s)}{s - q^2}\; ,
\label{OPE-hadron}
\end{eqnarray}
where $s_{min}$ is the kinematic limit, which usually corresponds to the square of the sum of current-quark masses of the hadron \cite{Wan:2020oxt,Wan:2020fsk}. $\rho^{X,\;OPE}_{J^{PC}}(s) = \text{Im} [\Pi^{X,\;OPE}_{J^{PC}}(s)] / \pi$ is the spectral density of the OPE side and contains the contributions of the condensates up to dimension 12 which can be expressed as:
\begin{eqnarray}
\rho^{OPE}(s) & = & \rho^{pert}(s) + \rho^{\langle \bar{q} q
\rangle}(s) +\rho^{\langle G^2 \rangle}(s) + \rho^{\langle \bar{q} G q \rangle}(s)
+ \rho^{\langle \bar{q} q \rangle^2}(s) + \rho^{\langle G^3 \rangle}(s) \nonumber\\
&+& \rho^{\langle \bar{q} q \rangle\langle \bar{q} G q \rangle}(s) +  \rho^{\langle \bar{q} q
\rangle^2\langle G^2 \rangle}(s) + \rho^{\langle \bar{q} G q \rangle^2}(s) +\rho^{\langle \bar{q} q \rangle^2\langle \bar{q} G q \rangle}(s)+ \rho^{\langle \bar{q} q
\rangle^4}(s)   \nonumber\\
&=& \sum_n c_n s^n  . \label{rho-OPE}
\end{eqnarray}
Here, $c_n$ is the coefficient of the $s^n$ term with corresponding vacuum condensates.

 To calculate the spectral density of the OPE side, Eq. (\ref{rho-OPE}), the full propagators $iS^q_{i j}(x)$ of a light quark ($q=u$, $d$, or $s$) are used:
\begin{eqnarray}
iS^q_{j k}(x) \! \! & = & \! \! \frac{i \delta_{j k} x\!\!\!\slash}{2 \pi^2
x^4} - \frac{\delta_{jk} m_q}{4 \pi^2 x^2} - \frac{i t^a_{j k} G^a_{\alpha\beta}}{32 \pi^2 x^2}(\sigma^{\alpha \beta} x\!\!\!\slash
+ x\!\!\!\slash \sigma^{\alpha \beta}) - \frac{\delta_{jk}}{12} \langle\bar{q} q \rangle + \frac{i\delta_{j k}
x\!\!\!\slash}{48} m_q \langle \bar{q}q \rangle - \frac{\delta_{j k} x^2}{192} \langle g_s \bar{q} \sigma \cdot G q \rangle \nonumber \\ &+& \frac{i \delta_{jk} x^2 x\!\!\!\slash}{1152} m_q \langle g_s \bar{q} \sigma \cdot G q \rangle - \frac{t^a_{j k} \sigma_{\alpha \beta}}{192}
\langle g_s \bar{q} \sigma \cdot G q \rangle
+ \frac{i t^a_{jk}}{768} (\sigma_{\alpha \beta} x \!\!\!\slash + x\!\!\!\slash \sigma_{\alpha \beta}) m_q \langle
g_s \bar{q} \sigma \cdot G q \rangle \;,
\end{eqnarray}
where, the vacuum condensates are clearly displayed. For more explanation on above propagator, readers may refer to Refs.~\cite{Wang:2013vex, Albuquerque:2013ija}.

Performing a Borel transform on both Eq. (\ref{hadron}) and Eq. (\ref{OPE-hadron}), and matching the OPE side with the phenomenological side of the correlation function $\Pi(q^2)$, one can finally obtain the mass of the hexaquark state,
\begin{eqnarray}
m^X_{J^{PC}}(s_0, M_B^2) = \sqrt{- \frac{L^{X}_{J^{PC},\;1}(s_0, M_B^2)}{L^{X}_{J^{PC},\;0}(s_0, M_B^2)}} \; . \label{mass-Eq}
\end{eqnarray}
Here $L_0$ and $L_1$ are respectively defined as
\begin{eqnarray}
L^{X}_{J^{PC},\;0}(s_0, M_B^2) =  \int_{s_{min}}^{s_0} d s \; \rho^{X,\;OPE}_{J^{PC}}(s) e^{-
s / M_B^2}  \;  \label{L0}
\end{eqnarray}
and
\begin{eqnarray}
L^{X}_{J^{PC},\;1}(s_0, M_B^2) =
\frac{\partial}{\partial{\frac{1}{M_B^2}}}{L^{X}_{J^{PC},\;0}(s_0, M_B^2)} \; .
\end{eqnarray}

\section{Numerical analysis}\label{Numerical}
In the numerical calculation of QCD sum rules, the input parameters are taken from \cite{Albuquerque:2013ija,Matheus:2006xi, Cui:2011fj, Narison:2002pw,P.Col,Tang:2019nwv,pdg}:
\begin{eqnarray}
\begin{aligned}
& \langle \bar{q} q \rangle = - (0.24 \pm 0.01)^3 \; \text{GeV}^3 \; ,
& &\langle \bar{s} s \rangle=(1.15\pm0.12)\langle \bar{q} q \rangle \; , \\
&\langle g_s^2 G^2 \rangle = (0.88\pm0.25) \; \text{GeV}^4  \; , & & \langle g_s^3 G^3 \rangle = (0.045\pm0.013) \;
\text{GeV}^6 \; ,\\
&\langle \bar{q} g_s \sigma \cdot G q \rangle = m_0^2 \langle
\bar{q} q \rangle \;, & &\langle \bar{s} g_s \sigma \cdot G s \rangle = m_0^2 \langle\bar{s} s \rangle \; ,\\
& m_0^2 = (0.8\pm0.1) \; \text{GeV}^2\; , & &m_s=(95\pm5)\; \text{MeV} \;,\\
&m_u=2.16^{+0.49}_{-0.26} \;\text{MeV}\; , & & m_d=4.67^{+0.48}_{-0.17}\; \text{MeV}\; ,
\end{aligned}
\end{eqnarray}
Note, a large strange quark condensate value is taken in numerical evaluation according to the recent lattice QCD calculation \cite{Davies:2018hmw}.

Furthermore, there exist two additional parameters, i.e., the continuum threshold $s_0$ and the Borel parameter $M_B^2$, introduced in establishing the sum rule, which can be fixed in light of the so-called standard procedures abiding by two criteria \cite{Shifman,Reinders:1984sr, P.Col}. The first one asks for the convergence of the OPE, which is to compare individual contributions with the overall magnitude on the OPE side, and then a reliable region for $M_B^2$ will be chosen to retain the convergence. The second criterion of QCD sum rules is the pole contribution (PC). As discussed in Refs. \cite{Chen:2014vha,Azizi:2019xla,Wang:2017sto}, the large power of $s$ in the spectral density suppresses the PC value; thus, the pole contribution will be chosen larger than $15\%$ for hexaquark states. The two criteria can be formulated as follows:
\begin{eqnarray}
  R^{X,\;OPE}_{J^{PC}} = \left| \frac{L_{J^{PC},\;0}^{X,\;c_0}(s_0, M_B^2)}{L_{J^{PC},\;0}^X(s_0, M_B^2)}\right|\, ,
\end{eqnarray}
\begin{eqnarray}
  R^{X,\;PC}_{J^{PC}} = \frac{L_{J^{PC},\;0}^X(s_0, M_B^2)}{L_{J^{PC},\;0}^X(\infty, M_B^2)} \; . \label{RatioPC}
\end{eqnarray}

Here, the superscript $c_0$ denotes the contribution given by the $c_0$ term. In addition, to find an optimal Borel window the contributions of $c_1$, $c_2$, and $c_3$ terms are also taken into account.

In order to determine a proper value for $s_0$, a similar analysis in Refs. \cite{Qiao:2013dda,Tang:2016pcf,Wan:2020oxt,Wan:2020fsk} will be carried out. Since the continuum threshold $s_0$ relates to the mass of the ground state by $\sqrt{s_0} \sim ( m^X + \delta )$, in which $\delta$ lies in the range $0.4-0.8$ GeV \cite{P.Col,Finazzo:2011he}, various $\sqrt{s_0}$ satisfying this constraint are taken into account. Among  these values, the one which yields an optimal window for Borel parameter $M_B^2$ should be selected out. That is, within the optimal window, the mass of hexaquark is somehow independent of the Borel parameter $M_B^2$ as much as possible. In practice, we will vary $\sqrt{s_0}$ by $0.1$ GeV, which obtained the lower and upper bounds hence the uncertainties of $\sqrt{s_0}$.

\begin{figure}
	\begin{center}
		\includegraphics[width=6.8cm]{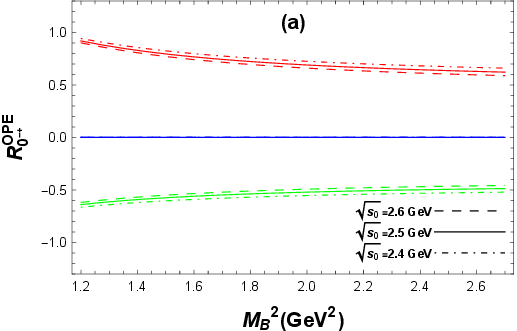}
		\includegraphics[width=6.8cm]{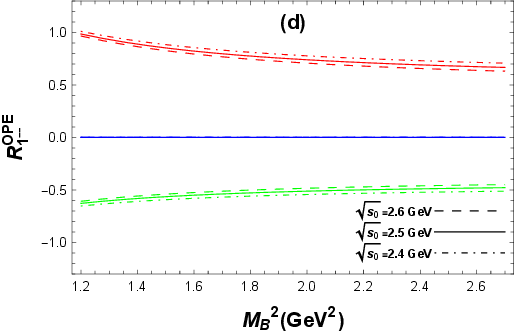}
		\includegraphics[width=6.8cm]{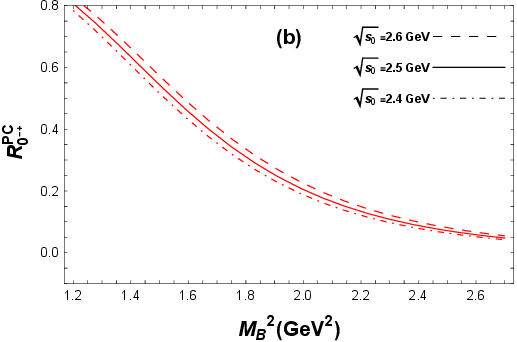}
		\includegraphics[width=6.8cm]{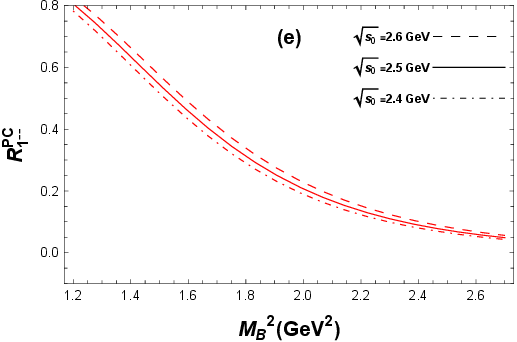}
		\includegraphics[width=6.8cm]{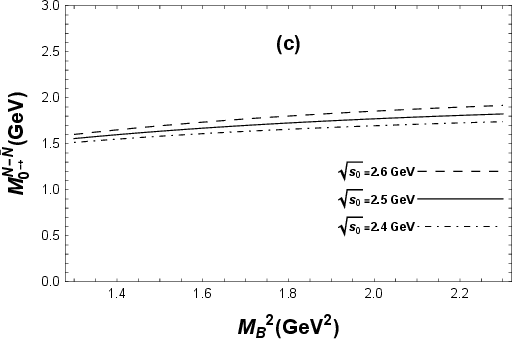}
		\includegraphics[width=6.8cm]{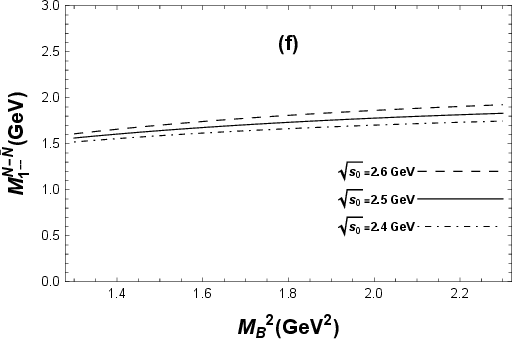}
		\caption{Figures for $N$-$\bar{N}$ baryonium states. \text{(a)} The ratio ${R_{0^{-+}}^{OPE}}$ state as functions of the Borel parameter $M_B^2$ for different values of $\sqrt{s_0}$, where blue lines, red lines, and green lines represent the contribution of $ c_0 $ terms, contributions of $c_1$ over $c_2$, and the contribution of $c_2$ over $c_3$, respectively. \text{(b)} The pole contribution ${R_{0^{-+}}^{PC}}$ as functions of the Borel parameter $M_B^2$ for different values of $\sqrt{s_0}$. \text{(c)} The mass $M_{0^{-+}}^{N\bar{N}}$ as a function of the Borel parameter $M_B^2$ for different values of $\sqrt{s_0}$. The same captions for \text{(d)}, \text{(e)} and \text{(f)} as in \text{(a)}, \text{(b)} and \text{(c)}, respectively, but for the $1^{--}$ state.} \label{figN}
	\end{center}
\end{figure}

With the above preparation the mass spectrum of baryonium in light sector will be  numerically evaluated. For $p$-$\bar{p}$ states, the ratios $R^{N\bar{N},\;OPE}_{0^{-+}}$ and $R^{N\bar{N},\;PC}_{0^{-+}}$ are shown as functions of Borel parameter $M_B^2$ in Fig. \ref{figN}(a) and Fig. \ref{figN}(b) with different values of $\sqrt{s_0}$, i.e., $2.4$, $2.5$, and $2.6$ GeV. Since the $c_0$ term will vanish in the chiral limit, we estimate the OPE convergence by inspecting the contributions of $c_1$ over $c_2$ and $c_2$ over $c_3$, which clearly indicate that higher dimensional terms yield relative small contributions, as shown in Figs. 1$ - $4. The dependence relationships between $m^{N\bar{N}}_{0^{-+}}$ and parameter $M_B^2$ are given in Fig. \ref{figN}(c). The optimal Borel window is found in the range $1.6 \le M_B^2 \le 2.2\; \text{GeV}^2$, and the mass $m^{N\bar{N}}_{0^{-+}}$ can be extracted as follows:
\begin{eqnarray}
m^{N\bar{N}}_{0^{-+}} &=& (1.81\pm 0.09)\; \text{GeV}.\label{m1}
\end{eqnarray}

The ratios $R^{N\bar{N},\;OPE}_{1^{--}}$ and $R^{N\bar{N},\;PC}_{1^{--}}$ are presented in Fig. \ref{figN}(d) and Fig. \ref{figN}(e) with different values of $\sqrt{s_0}$, i.e., $2.4$, $2.5$, and $2.6$ GeV, and the relationships between $m^{N\bar{N}}_{1^{--}}$ and parameter $M_B^2$ are displayed in Fig. \ref{figN}(f). The optimal Borel window is found in $1.6 \le M_B^2 \le 2.2\; \text{GeV}^2$, and the mass $m^{N\bar{N}}_{1^{--}}$ can be evaluated as follows:
\begin{eqnarray}
m^{N\bar{N}}_{1^{--}} &=& (1.82\pm 0.10)\; \text{GeV}.\label{m2}
\end{eqnarray}

\begin{figure}
	\begin{center}
		\includegraphics[width=6.8cm]{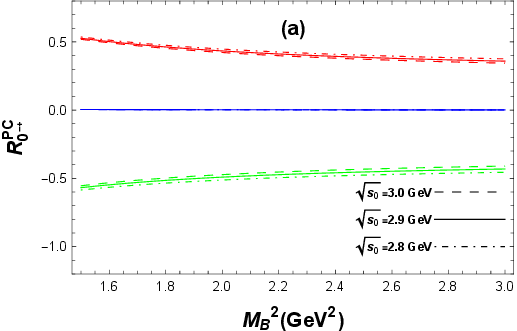}
		\includegraphics[width=6.8cm]{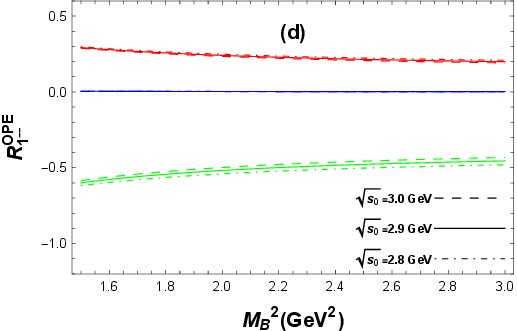}
		\includegraphics[width=6.8cm]{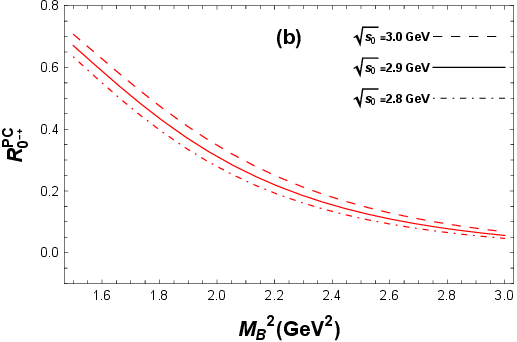}
		\includegraphics[width=6.8cm]{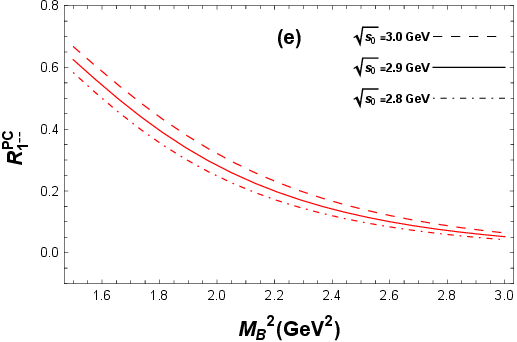}
		\includegraphics[width=6.8cm]{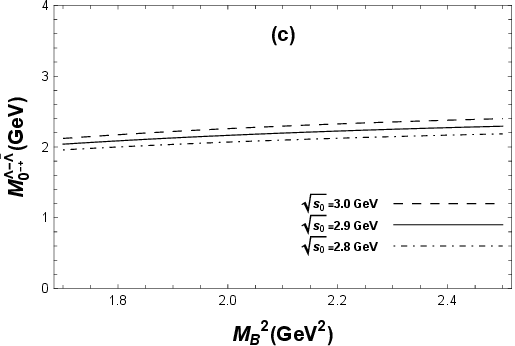}
		\includegraphics[width=6.8cm]{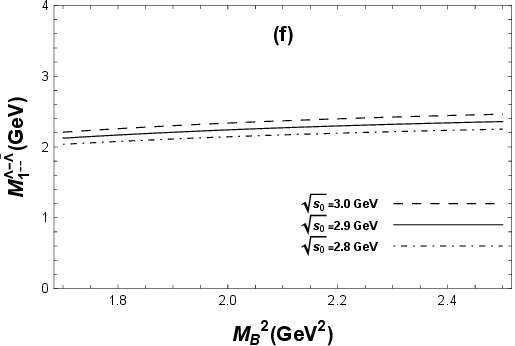}
		\caption{The same caption as in Fig. \ref{figN}, but for the $\Lambda$-$\bar{\Lambda}$ states.} \label{figL}
	\end{center}
\end{figure}

For $\Lambda$-$\bar{\Lambda}$ states, we show the ratios $R^{\Lambda\bar{\Lambda},\;OPE}_{0^{-+}}$ and $R^{\Lambda\bar{\Lambda},\;PC}_{0^{-+}}$ in Fig. \ref{figL}(a) and Fig. \ref{figL}(b) with different values of $\sqrt{s_0}$, i.e., $2.8$, $2.9$, and $3.0$ GeV, and the relationships between $m^{\Lambda\bar{\Lambda}}_{0^{-+}}$ and parameter $M_B^2$ are given in Fig. \ref{figL}(c). The optimal Borel window is found in $1.9 \le M_B^2 \le 2.5\; \text{GeV}^2$, and the mass $m^{\Lambda\bar{\Lambda}}_{0^{-+}}$ can be obtained as follows:
\begin{eqnarray}
m^{\Lambda\bar{\Lambda}}_{0^{-+}} &=& (2.27\pm 0.13)\; \text{GeV}.\label{m3}
\end{eqnarray}

The ratios $R^{\Lambda\bar{\Lambda},\;OPE}_{1^{--}}$ and $R^{\Lambda\bar{\Lambda},\;PC}_{1^{--}}$ are shown in Fig. \ref{figL}(d) and Fig. \ref{figL}(e), where $\sqrt{s_0}=2.8$, $2.9$, and $3.0$ GeV, respectively, and we display the relationships between $m^{\Lambda\bar{\Lambda}}_{1^{--}}$ and parameter $M_B^2$ in Fig. \ref{figL}(f). The optimal Borel window is found in $1.9 \le M_B^2 \le 2.5\; \text{GeV}^2$, and the mass $m^{\Lambda\bar{\Lambda}}_{1^{--}}$ can be estimated as follows:
\begin{eqnarray}
m^{\Lambda\bar{\Lambda}}_{1^{--}} &=& (2.34\pm 0.12)\; \text{GeV}.\label{m4}
\end{eqnarray}

\begin{figure}[h]
	\begin{center}
		\includegraphics[width=6.8cm]{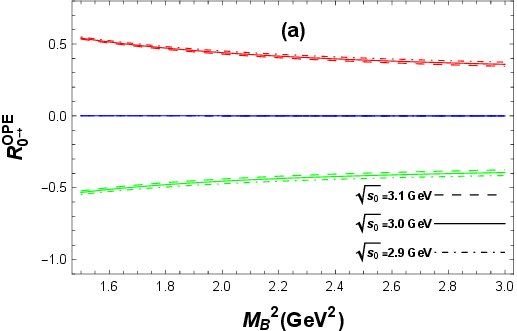}
		\includegraphics[width=6.8cm]{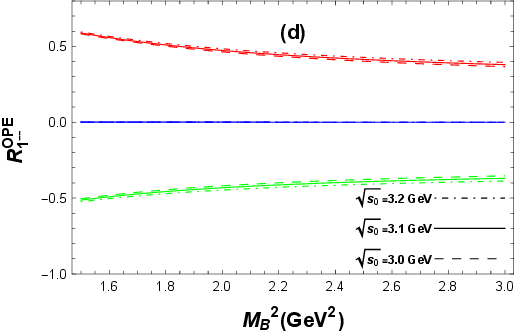}
		\includegraphics[width=6.8cm]{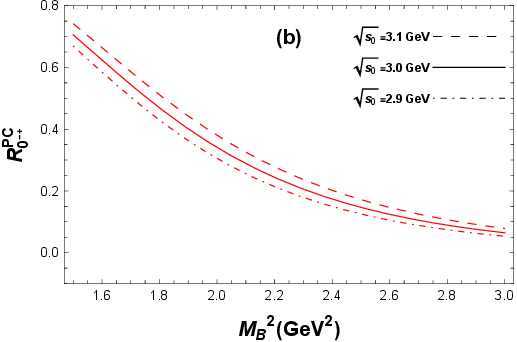}
		\includegraphics[width=6.8cm]{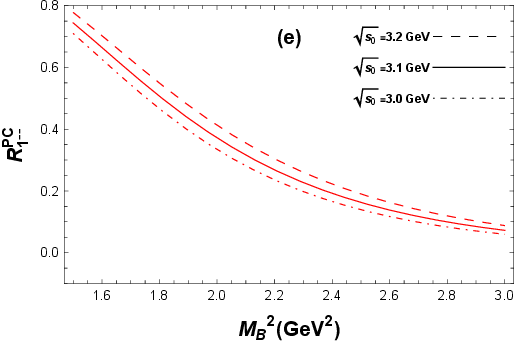}
		\includegraphics[width=6.8cm]{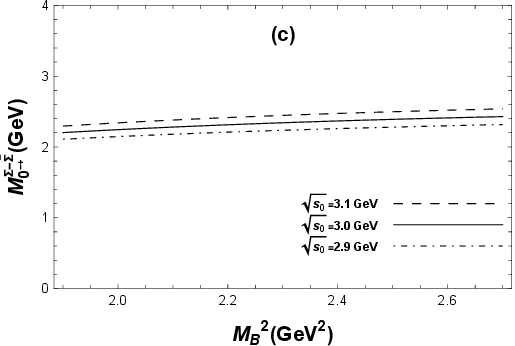}
		\includegraphics[width=6.8cm]{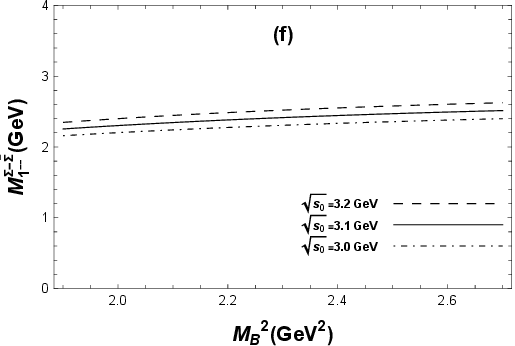}
		\caption{The same caption as in Fig. \ref{figN}, but for the $\Sigma$-$\bar{\Sigma}$ states.} \label{figS}
	\end{center}
\end{figure}

For the $\Sigma$-$\bar{\Sigma}$ states, the ratios $R^{\Sigma\bar{\Sigma},\;OPE}_{0^{-+}}$ and $R^{\Sigma\bar{\Sigma},\;PC}_{0^{-+}}$ can be found in Fig. \ref{figS}(a) and Fig. \ref{figS}(b) with different values of $\sqrt{s_0}$, i.e., $2.9$, $3.0$, and $3.1$ GeV, and the relationships between $m^{\Sigma\bar{\Sigma}}_{0^{-+}}$ and parameter $M_B^2$ are presented in Fig. \ref{figS}(c). The optimal Borel window is found in $2.0 \le M_B^2 \le 2.6\; \text{GeV}^2$, and the mass $m^{\Sigma\bar{\Sigma}}_{0^{-+}}$ can be acquired as follows:
\begin{eqnarray}
m^{\Sigma\bar{\Sigma}}_{0^{-+}} &=& (2.39\pm 0.13)\; \text{GeV}.\label{m5}
\end{eqnarray}

We present the ratios $R^{\Sigma\bar{\Sigma},\;OPE}_{1^{--}}$ and $R^{\Sigma\bar{\Sigma},\;PC}_{1^{--}}$ in Fig. \ref{figS}(d) and Fig. \ref{figS}(e) with different values of $\sqrt{s_0}$, i.e., $3.0$, $3.1$, and $3.2$ GeV, and the relationships between $m^{\Sigma\bar{\Sigma}}_{1^{--}}$ and parameter $M_B^2$ are shown in Fig. \ref{figS}(f). The optimal Borel window is found in $2.0 \le M_B^2 \le 2.6\; \text{GeV}^2$, and the mass $m^{\Sigma\bar{\Sigma}}_{1^{--}}$ can be extracted as follows:
\begin{eqnarray}
m^{\Sigma\bar{\Sigma}}_{1^{--}} &=& (2.48\pm 0.12)\; \text{GeV}.\label{m6}
\end{eqnarray}

\begin{figure}
	\begin{center}
		\includegraphics[width=6.8cm]{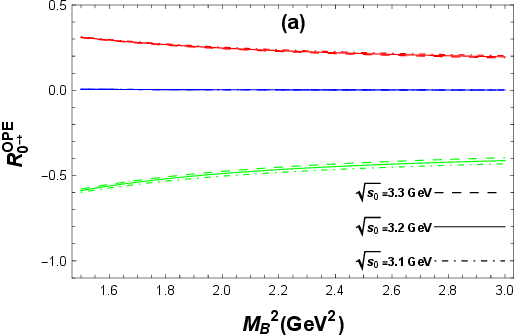}
		\includegraphics[width=6.8cm]{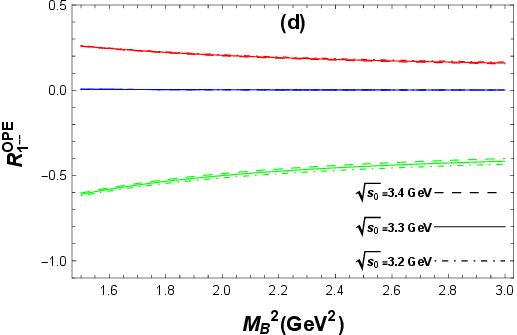}
		\includegraphics[width=6.8cm]{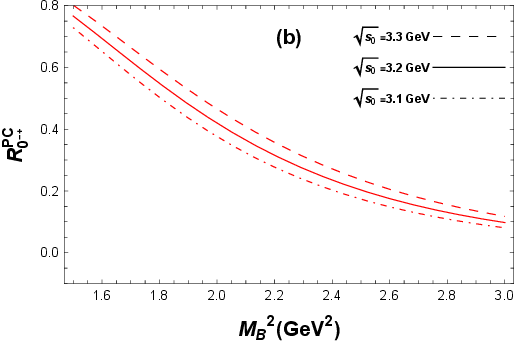}
		\includegraphics[width=6.8cm]{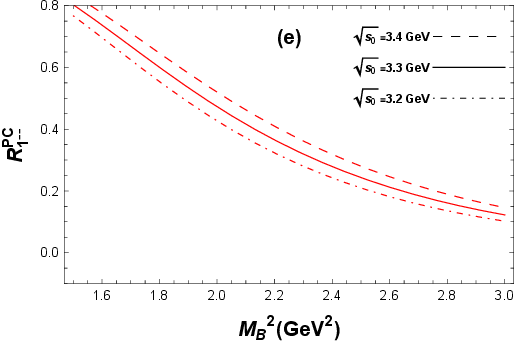}
		\includegraphics[width=6.8cm]{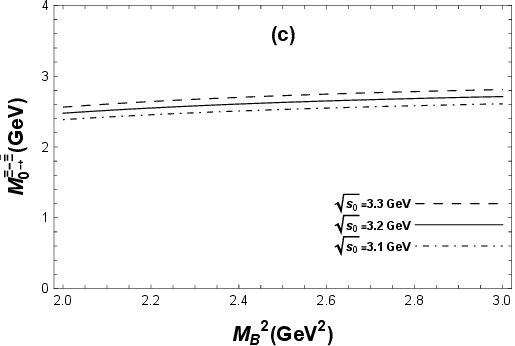}
		\includegraphics[width=6.8cm]{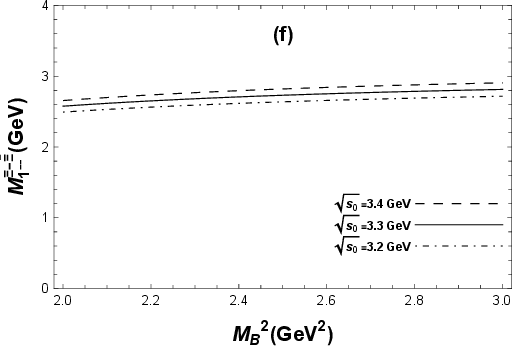}
		\caption{The same caption as in Fig. \ref{figN}, but for the $\Xi$-$\bar{\Xi}$ states.} \label{figX}
	\end{center}
\end{figure}

For the $\Xi$-$\bar{\Xi}$ states, the ratios $R^{\Xi\bar{\Xi},\;OPE}_{0^{-+}}$ and $R^{\Xi\bar{\Xi},\;PC}_{0^{-+}}$ are shown in Fig. \ref{figX}(a) and Fig. \ref{figX}(b) with $\sqrt{s_0}=3.1$, $3.2$, and $3.3$ GeV, respectively, and the relationships between $m^{\Xi\bar{\Xi}}_{0^{-+}}$ and parameter $M_B^2$ are given in Fig. \ref{figX}(c). The optimal Borel window is found in $2.1 \le M_B^2 \le 2.8\; \text{GeV}^2$, and the mass $m^{\Xi\bar{\Xi}}_{0^{-+}}$ can be obtained as follows:
\begin{eqnarray}
m^{\Xi\bar{\Xi}}_{0^{-+}} &=& (2.67\pm 0.12)\; \text{GeV}.\label{m7}
\end{eqnarray}

The ratios $R^{\Xi\bar{\Xi},\;OPE}_{1^{--}}$ and $R^{\Xi\bar{\Xi},\;PC}_{1^{--}}$ are shown in Fig. \ref{figX}(d) and Fig. \ref{figX}(e)with different values of $\sqrt{s_0}$, i.e., $3.2$, $3.3$, and $3.4$ GeV, and the relationships between $m^{\Xi\bar{\Xi}}_{1^{--}}$ and parameter $M_B^2$ are given in Fig. \ref{figX}(f). The optimal Borel window is found in $2.1 \le M_B^2 \le 2.9\; \text{GeV}^2$, and the mass $m^{\Xi\bar{\Xi}}_{1^{--}}$ can be evaluated as follows:
\begin{eqnarray}
m^{\Xi\bar{\Xi}}_{1^{--}} &=& (2.79\pm 0.11)\; \text{GeV}.\label{m8}
\end{eqnarray}

The errors of the results, (\ref{m1})$-$(\ref{m8}), mainly stem from the uncertainties in quark masses, condensates, Borel parameter $M_B^2$ and threshold parameter $\sqrt{s_0}$. The nonperturbative contributions by dimensions higher than 12 are also evaluated, and it is less than $10\%$ of dimension 12, so the convergence of OPE makes the high-dimension condensates contributions negligible.

We also analyze the $\mathscr{B}$-$\bar{\mathscr{B}}$ states with the situations of $0^{++}$ and $1^{++}$, and find that no matter what values of $M_B^2$ and $\sqrt{s_0}$ take, no optimal window for stable plateaus exists. That means the currents in Eqs. (\ref{Ja0++}) and (\ref{Ja1++}) do not support the corresponding $\mathscr{B}$-$\bar{\mathscr{B}}$ hexaquark molecular states.

For the convenience of reference, a collection of continuum thresholds, Borel parameters, and predicted masses of hexaquark states are tabulated in Table \ref{mass_baryonium}.
\begin{table}
\begin{center}
\renewcommand\tabcolsep{10pt}
\caption{The continuum thresholds, Borel parameters, and predicted masses of hexaquark molecular states.}\label{mass_baryonium}
\begin{tabular}{ccccc}\hline\hline
$\mathscr{B}$-$\bar{\mathscr{B}}$  &$J^{PC}$   & $\sqrt{s_0}\;(\text{GeV})$     &$M_B^2\;(\text{GeV}^2)$ &$M^X\;(\text{GeV})$       \\ \hline
$N-\bar{N}$                                         &$0^{-+}$    & $2.5\pm0.1$                             &$1.6-2.2$                             &$1.81\pm0.09$                \\
                                                              &$1^{--}$     & $2.5\pm0.1$                             &$1.6-2.2$                              &$1.82\pm0.10$                   \\
                                                              &$0^{++}$   &     $...$                                        &$...$                                        &$...$                                     \\
                                                              &$1^{++}$   &     $...$                                        &$...$                                        &$...$                                    \\ \hline
$\Lambda-\bar{\Lambda}$                 &$0^{-+}$    & $2.9\pm0.1$                             &$1.9-2.5$                              &$2.27\pm0.13$               \\
                                                              &$1^{--}$     & $2.9\pm0.1$                            &$1.9-2.5$                               &$2.34\pm0.12$                \\
                                                              &$0^{++}$   &     $...$                                         &$...$                                        &$...$                                    \\
                                                              &$1^{++}$   &     $...$                                        &$...$                                        &$...$                                      \\ \hline
$\Sigma-\bar{\Sigma}$                       &$0^{-+}$    & $3.0\pm0.1$                             &$2.0-2.6$                                &$2.39\pm0.13$                \\
                                                              &$1^{--}$     & $3.1\pm0.1$                            & $2.0-2.6$                              &$2.48\pm0.12$                  \\
                                                              &$0^{++}$   &     $...$                                       &$...$                                        &$...$                                    \\
                                                              &$1^{++}$   &     $...$                                       &$...$                                        &$...$                                    \\ \hline
$\Xi-\bar{\Xi}$                                  &$0^{-+}$    & $3.2\pm0.1$                             &$2.1-2.8$                               &$2.67\pm0.12$                 \\
                                                              &$1^{--}$     & $3.3\pm0.1$                          &$2.1-2.9$                                &$2.79\pm0.11$                  \\
                                                              &$0^{++}$   &     $...$                                      &$...$                                        &$...$                                   \\
                                                              &$1^{++}$   &     $...$                                      &$...$                                        &$...$                                      \\
\hline
 \hline
\end{tabular}
\end{center}
\end{table}

\section{Decay analyses}\label{Decay}
To finally ascertain these light baryonium states, the straightforward procedure is to reconstruct them from their decay products, though the detailed characters still ask for more investigation. In our evaluation, the masses of $\Lambda$-$\bar{\Lambda}$, $\Sigma$-$\bar{\Sigma}$, and $\Xi$-$\bar{\Xi}$ states are above the  threshold of their respective $\mathscr{B}\bar{\mathscr{B}}$ dibaryons, so the $\Lambda\bar{\Lambda}$, $\Sigma\bar{\Sigma}$, and $\Xi$$\bar{\Xi}$ decay channel will be  the primary decay mode, respectively. On the other hand, the masses of $p$-$\bar{p}$ states are below the threshold of $p\bar{p}$ dibaryon, so the $p\bar{p}$ decay channel should be forbidden. The typical decay modes of the light baryonium for different quantum numbers are given in Table \ref{decay-mode}, and these processes are expected to be measurable in the running experiments like the BESIII, BELLEII, and LHCb.

\begin{table}
\begin{center}\caption{Typical decay modes of the light baryonium for each quantum number.}\label{decay-mode}
\renewcommand\tabcolsep{8.0pt}
\begin{tabular}{cccccccc}
\hline
\hline
$J^{PC}$  &$N$-$\bar{N}$         & $\Lambda$-$\bar{\Lambda}$      &$\Sigma$-$\bar{\Sigma}$      &$\Xi$-$\bar{\Xi}$   \\
\hline
$0^{-+}$   &$\pi\pi\pi$                 &$\Lambda\bar{\Lambda}$             &$\Sigma\bar{\Sigma}$            &$\Xi\bar{\Xi}$\;\;\;$\eta K K$ \\
		       &$\pi\pi\eta$                &$\pi\pi\eta$                                     &$\pi\pi\eta$                              &$\eta\eta\pi$\;\;\;$\eta^\prime K K$        \\
		       &$\pi\pi\eta^\prime$    &$\pi\pi\eta^\prime$                         &$\pi\pi\eta^\prime$                   &$\eta^\prime\eta^\prime\pi$ \\
			  &						       &$\pi K K$  	                                  &$\pi K K$                                 &$\eta\eta^\prime\pi$      \\
\hline
$1^{--}$ &$\pi\pi\omega$           &$\Lambda\bar{\Lambda}$              &$\Sigma\bar{\Sigma}$              &$\Xi\bar{\Xi}$\;\;\;$\omega K K$  \\
               &$\pi\pi\phi$                &$\pi\pi\omega$                                 &$\pi\pi\omega$				      &$\phi\phi\pi$\;\;\;$\phi K K$             \\
                &						    &$\pi\pi\phi$                                        &$\pi\pi\phi$                              &$\omega\omega\pi$   \\
               &						    &$\pi K^\ast K^\ast$			         	&$\pi K^\ast K^\ast$                	&$\omega\phi\pi$ \\
\hline
\hline
\end{tabular}
\end{center}
\end{table}

\section{Conclusions}

In summary, we investigate the light baryonium states in molecular configuration with quantum numbers of $J^{PC}=0^{-+}$, $1^{--}$, $0^{++}$, and $1^{++}$ in the framework of QCD sum rules. Our results suggest that there exist eight possible light baryonium states, i.e., $p$-$\bar{p}$, $\Lambda$-$\bar{\Lambda}$, $\Sigma$-$\bar{\Sigma}$, and $\Xi$-$\bar{\Xi}$ with quantum numbers of $0^{-+}$  and $1^{--}$, and their masses are tabulated in Table \ref{mass_baryonium}. According to our evaluation, the masses of $\Lambda$-$\bar{\Lambda}$, $\Sigma$-$\bar{\Sigma}$, and $\Xi$-$\bar{\Xi}$ states are above their corresponding dibaryon thresholds, while the masses of $p$-$\bar{p}$ states are not. Moreover, the primary and potential decay modes of these light baryonium states are analyzed, which might serve as a guide for experimental exploration in BESIII, BELLEII, or LHCb.

\vspace{.5cm} {\bf Acknowledgments} \vspace{.5cm}

This work was supported in part by the National Key Research and Development Program of China under Contract No. 2020YFA0406400, and the National Natural Science Foundation of China (NSFC) under the Grants No. 11975236 and No. 11635009.


\begin{widetext}
\appendix

\section{The spectral densities $\mathscr{B}$-$\bar{\mathscr{B}}$ hexaquark states}

\subsection{The spectral densities of $0^{-+}$ $p$-$\bar{p}$ hexaquark states}

The $0^{-+}$ $p$-$\bar{p}$ hexaquark state spectral densities on the OPE side can be expressed as:
\begin{eqnarray}
\Pi_{0^{-+}}^{ p\bar{p}}(s_0,M_B^2)=\int_{s_{min}}^{s_0}\;e^{-s/M_{B}^2}\;\sum_n\;c_n\;s^n\;ds + \Pi^{sum}(M_B^2)\;.
\end{eqnarray}

Here,
\begin{eqnarray}
c_7&=&\frac{1}{2^{23}\times 3^2 \times 5^2 \times 7^2 \; \pi ^{10}}\;,\\
c_6&=&\frac{m_u \left(2 m_d+m_u\right) }{2^{20}\times 3^2 \times 5^2 \times 7 \; \pi ^{10}}\;,\\
c_5&=&\frac{m_d \left(m_d+2 m_u\right) m_u^2}{2^{19}\times 3 \times 5^2  \, \pi ^{10}}-\frac{\left\langle \bar{q}q \right\rangle \left(m_d+2 m_u\right)}{2^{17}\times 3^2 \times 5^2  \, \pi ^8}+\frac{\left\langle GG \right\rangle}{2^{22}\times 3^2 \times 5^2  \, \pi ^{10}} \;,\\
c_4&=&\frac{\left\langle GG \right\rangle \left(m_d+m_u\right) m_u}{2^{21}\times 3 \times 5  \, \pi ^{10}}-\frac{\left\langle \bar{q}q \right\rangle \left(m_d^2+4 m_u m_d+m_u^2\right) m_u}{2^{16}\times 3 \times 5  \, \pi ^8} \nonumber\\
&+&\frac{m_d^2 m_u^4}{2^{18}\times 5  \, \pi ^{10}}+\frac{\left\langle \bar{q}Gq \right\rangle \left(m_d+2 m_u\right)}{2^{17}\times 3 \times 5  \, \pi ^8}+\frac{\left\langle \bar{q}q \right\rangle^2}{2^{14}\times 3 \times 5  \, \pi ^6}\;,\\
c_3&=&\frac{\left\langle GG \right\rangle m_d m_u^3}{2^{19}\times 3  \, \pi ^{10}}+\frac{\left\langle \bar{q}Gq \right\rangle \left(m_d^2+4 m_u m_d+m_u^2\right) m_u}{2^{15}\times 3  \, \pi ^8}-\left\langle \bar{q}q \right\rangle \bigg[\frac{m_d \left(2 m_d+m_u\right) m_u^3}{2^{14}\times 3  \, \pi ^8} \nonumber\\
&+&\frac{\left\langle \bar{q}Gq \right\rangle}{2^{13}\times 3  \, \pi ^6}+\frac{\left\langle GG \right\rangle \left(m_d+3 m_u\right)}{2^{18}\times 3^2  \, \pi ^8}\bigg]+\frac{\left\langle \bar{q}q \right\rangle^2 \left(m_d^2+24 m_u m_d+14 m_u^2\right)}{2^{14}\times 3^2  \, \pi ^6}\;,\\
c_2&=&-\frac{\left(m_d+2 m_u\right) \left\langle \bar{q}q \right\rangle^3}{2^{7}\times 3^2  \, \pi ^4}+\bigg[\frac{\left(11 m_d^2+18 m_u m_d+m_u^2\right) m_u^2}{2^{12}\times 3 \, \pi ^6}+\frac{\left\langle GG \right\rangle}{2^{12}\times 3^2 \, \pi ^6}\bigg] \left\langle \bar{q}q \right\rangle^2\nonumber\\
&+&\bigg[-\frac{\left\langle GG \right\rangle \left(3 m_d+m_u\right) m_u^2}{2^{16}\times 3 \, \pi ^8}-\frac{\left\langle \bar{q}Gq \right\rangle \left(m_d^2+14 m_u m_d+9 m_u^2\right)}{2^{12}\times 3 \, \pi ^6}\bigg] \left\langle \bar{q}q \right\rangle\nonumber\\
&+&\frac{\left\langle \bar{q}Gq \right\rangle m_d m_u^3 \left(2 m_d+m_u\right)}{2^{13} \, \pi ^8}+\frac{\left\langle GG \right\rangle \left\langle \bar{q}Gq \right\rangle \left(m_d+3 m_u\right)}{2^{17}\times 3 \, \pi ^8}+\frac{\left\langle \bar{q}Gq \right\rangle^2}{2^{14} \, \pi ^6}\;,\\
c_1&=&\frac{\left\langle \bar{q}q \right\rangle^4}{2^{5}\times 3 \, \pi ^2}-\frac{m_u \left(3 m_d^2+11 m_u m_d+3 m_u^2\right) \left\langle \bar{q}q \right\rangle^3}{2^{8}\times 3 \, \pi ^4}\nonumber\\
&+&\left[\frac{3 m_d^2 m_u^4}{2^{11} \, \pi ^6}+\frac{7 \left\langle GG \right\rangle \left(m_d+m_u\right) m_u}{2^{14}\times 3  \, \pi ^6}+\frac{\left\langle \bar{q}Gq \right\rangle \left(m_d+2 m_u\right)}{2^{6}\times 3 \, \pi ^4}\right] \left\langle \bar{q}q \right\rangle^2\nonumber\\
&+&\left[-\frac{\left\langle \bar{q}Gq \right\rangle \left(7 m_d^2+10 m_u m_d+m_u^2\right) m_u^2}{2^{11} \, \pi ^6}-\frac{\left\langle GG \right\rangle \left\langle \bar{q}Gq \right\rangle}{2^{12}\times 3 \, \pi ^6}\right] \left\langle \bar{q}q \right\rangle\nonumber\\
&+&\frac{\left\langle GG \right\rangle \left\langle \bar{q}Gq \right\rangle m_u^2 \left(3 m_d+m_u\right)}{2^{16} \, \pi ^8}+\frac{\left\langle \bar{q}Gq \right\rangle^2 \left(m_d^2+10 m_u m_d+7 m_u^2\right)}{2^{13} \, \pi ^6}\;,\\
c_0&=&\left\langle \bar{q}q \right\rangle^2 \left[\frac{\left\langle GG \right\rangle m_d m_u^3}{2^{12}\times 3 \, \pi ^6}+\frac{\left\langle \bar{q}Gq \right\rangle \left(2 m_d^2+7 m_u m_d+2 m_u^2\right) m_u}{2^{8} \, \pi ^4}\right]\nonumber\\
&+&\frac{\left\langle \bar{q}Gq \right\rangle^2 m_u^2 \left(6 m_d^2+8 m_u m_d+m_u^2\right)}{2^{12} \, \pi ^6}+\frac{\left\langle \bar{q}q \right\rangle^4 \left(m_d^2+18 m_u m_d+11 m_u^2\right)}{2^{6}\times 3^2 \, \pi ^2}\nonumber\\
&-&\frac{m_d m_u^3 \left\langle \bar{q}q \right\rangle^3 \left(2 m_d+m_u\right)}{2^{7}\times 3 \, \pi ^4}-\frac{m_u \left\langle \bar{q}q \right\rangle \left\langle GG \right\rangle \left\langle \bar{q}Gq \right\rangle \left(m_d+m_u\right)}{2^{11}\times 3 \, \pi ^6}\;,
\end{eqnarray}
and $\Pi^{sum}$ is the sum of those contributions in the correlation function that have no imaginary part but are nontrivial after the Borel transformation, and
\begin{eqnarray}
\Pi^{sum}(M_B^2)=\frac{\left\langle \bar{q}q \right\rangle^4 m_u^2 \left(24 m_d m_u+14 m_d^2+m_u^2\right)}{2^{8}\times 3^2 \, \pi ^2}+\frac{\left\langle \bar{q}q \right\rangle^2 \left\langle \bar{q}Gq \right\rangle m_d m_u^3 \left(2 m_d+m_u\right)}{2^{8}\times 3 \, \pi ^4}\;.
\end{eqnarray}

\subsection{The spectral densities of $0^{++}$ $p$-$\bar{p}$ hexaquark states}
The $0^{++}$ $p$-$\bar{p}$ hexaquark state spectral densities on the OPE side can be expressed as:
\begin{eqnarray}
c_7&=&-\frac{1}{2^{23}\times 3^2\times 5^2\times 7^2\, \pi ^{10}}\;,\\
c_6&=&\frac{ m_u \left(m_u-2 m_d\right)}{2^{20}\times 3^2\times 5^2\times 7^1\, \pi ^{10}}\;,\\
c_5&=&\frac{\left\langle \bar{q}q \right\rangle \left(m_d-2 m_u\right)}{2^{17}\times 3^2\times 5^2\, \pi ^8}+\frac{m_u^2 \left(2 m_d m_u-m_d^2\right)}{2^{19}\times 3\times 5^2\, \pi ^{10}}-\frac{\left\langle GG \right\rangle^{2}}{2^{22}\times 3^2\times 5^2\, \pi ^{10}}\;,\\
c_4&=&\frac{\left\langle \bar{q}q \right\rangle m_u \left(m_d^2-m_u \left(6 m_d+m_u\right)\right)}{2^{16}\times 3\times 5\, \pi ^8}+\frac{\left\langle GG \right\rangle^{2} m_u \left(m_u-m_d\right)}{2^{21}\times 3\times 5\, \pi ^{10}}\nonumber\\
&-&\frac{\left\langle \bar{q}Gq \right\rangle m_d}{2^{17}\times 3\times 5\, \pi ^8}+\frac{9 m_d^2 m_u^4-16 \pi ^4 \left\langle \bar{q}q \right\rangle^{2}}{2^{18}\times 3^2\times 5\, \pi ^{10}}\;,\\
c_3&=&\left\langle \bar{q}q \right\rangle \left[\frac{\left\langle GG \right\rangle^{2} \left(m_d-5 m_u\right)}{2^{18}\times 3^2\, \pi ^8}-\frac{m_u^3 \left(m_d m_u+4 m_d^2\right)}{2^{14}\times 3 \, \pi ^8}\right]\nonumber\\
&+&\frac{\left\langle GG \right\rangle^{2} m_d m_u^3}{2^{19}\times 3 \, \pi ^{10}}+\frac{\left\langle \bar{q}Gq \right\rangle m_u \left(2 m_d m_u-m_d^2+m_u^2\right)}{2^{15}\times 3 \, \pi ^8}\nonumber\\
&+&\frac{2 \left\langle \bar{q}q \right\rangle^{2} m_u \left(4 m_d+9 m_u\right)+\left\langle \bar{q}q \right\rangle^{2} \left(-m_d^2\right)+2 \left\langle \bar{q}q \right\rangle \left\langle \bar{q}Gq \right\rangle }{2^{14}\times 3^2 \, \pi ^6}\;,\\
c_2&=&-\frac{\left\langle \bar{q}q \right\rangle \left\langle GG \right\rangle^{2} m_u^2 \left(7 m_d+m_u\right)}{2^{16}\times 3 \, \pi ^8}+\frac{\left\langle GG \right\rangle^{2} \left\langle \bar{q}Gq \right\rangle \left(m_u-m_d\right)}{2^{17}\times 3 \, \pi ^8}+\frac{\left\langle \bar{q}Gq \right\rangle m_u^3 \left(m_d m_u+2 m_d^2\right)}{2^{13}\, \pi ^8}\nonumber\\
&+&\frac{1}{2^{12}\times 3 \, \pi ^6}\bigg[m_u^2 \left(15 \left\langle \bar{q}q \right\rangle^{2} m_d^2-11 \left\langle \bar{q}q \right\rangle \left\langle \bar{q}Gq \right\rangle \right)+m_d \left(22 \left\langle \bar{q}q \right\rangle^{2} m_u^3-6 \left\langle \bar{q}q \right\rangle \left\langle \bar{q}Gq \right\rangle  m_u\right)\nonumber\\
&+&\left\langle \bar{q}q \right\rangle \left\langle \bar{q}Gq \right\rangle  m_d^2+\left\langle \bar{q}q \right\rangle^{2} m_u^4\bigg]-\frac{\left\langle \bar{q}q \right\rangle^3 m_u}{2^{7}\times 3 \, \pi ^4}-\frac{\left\langle \bar{q}Gq \right\rangle^2}{2^{14}\times 3 \, \pi ^6}\;,\\
c_1&=&-\frac{\left\langle \bar{q}q \right\rangle^3 m_u \left(7 m_d m_u+m_d^2+m_u^2\right)}{2^{8}\, \pi ^4}+\left\langle \bar{q}q \right\rangle^2 \left[\frac{3 \left\langle GG \right\rangle^{2} m_u \left(m_d+m_u\right)}{2^{14}\, \pi ^6}+\frac{\left\langle \bar{q}Gq \right\rangle m_u}{2^{5} \times 3\, \pi ^4}\right]\nonumber\\
&+&\frac{\left\langle GG \right\rangle^{2} \left\langle \bar{q}Gq \right\rangle m_u^2 \left(3 m_d+m_u\right)}{2^{16}\, \pi ^8}-\frac{\left\langle \bar{q}Gq \right\rangle^2 \left(-2 m_d m_u+m_d^2-5 m_u^2\right)}{2^{13}\, \pi ^6}+\frac{\left\langle \bar{q}q \right\rangle^4}{2^{5}\times 3^2\, \pi ^2}\nonumber\\
&+&\frac{m_u^2 \left[m_u^2 (\left\langle \bar{q}q \right\rangle^{2} m_d^2-\left\langle \bar{q}q \right\rangle \left\langle \bar{q}Gq \right\rangle )-14 \left\langle \bar{q}q \right\rangle \left\langle \bar{q}Gq \right\rangle  m_d m_u-11 \left\langle \bar{q}q \right\rangle \left\langle \bar{q}Gq \right\rangle  m_d^2\right]}{2^{11}\, \pi ^6}\;,\\
c_0&=&\frac{\left\langle \bar{q}q \right\rangle^4 \left(22 m_d m_u+m_d^2+15 m_u^2\right)}{2^{6}\times 3^2 \, \pi ^2}-\frac{\left\langle \bar{q}q \right\rangle^3 m_d^2 m_u^3}{2^{7} \, \pi ^4}-\frac{\left\langle \bar{q}q \right\rangle \left\langle GG \right\rangle^{2} \left\langle \bar{q}Gq \right\rangle m_u \left(m_d+m_u\right)}{2^{12} \, \pi ^6}\nonumber\\
&+&\frac{\left\langle \bar{q}q \right\rangle^2 \left\langle \bar{q}Gq \right\rangle m_u \left(31 m_d m_u+7 m_d^2+7 m_u^2\right)}{2^{8}\times 3 \, \pi ^4}+\frac{\left\langle \bar{q}Gq \right\rangle^2 m_u^2 \left(8 m_d m_u+6 m_d^2+m_u^2\right)}{2^{12} \, \pi ^6}\;,\nonumber\\
\end{eqnarray}
and
\begin{eqnarray}
\Pi^{sum}(M_B^2)=\frac{\left\langle \bar{q}q \right\rangle^{2} \left\langle \bar{q}Gq \right\rangle \left(m_d^2 m_u^3\right)}{2^{8}\times 3 \, \pi ^4}+\left\langle \bar{q}q \right\rangle^4 \left(\frac{m_d m_u^3}{2^{5}\times 3^2\, \pi ^2}++\frac{m_d^2 m_u^2}{2^{7} \, \pi ^2}-\frac{m_u^4}{2^{8}\times 3^2 \, \pi ^2}\right)\;.
\end{eqnarray}

\subsection{The spectral densities of $1^{--}$ $p$-$\bar{p}$ hexaquark states}
The $1^{--}$ $p$-$\bar{p}$ hexaquark state spectral densities on the OPE side can be expressed as:
\begin{eqnarray}
c_7&=&-\frac{1}{2^{20}\times 3^4\times 5^2 \times 7^2 \, \pi ^{10}}\;,\\
c_6&=&-\frac{ m_u \left(7 m_d+4 m_u\right)}{2^{22}\times 3^2 \times 5^2 \times 7 \, \pi ^{10}}\;,\\
c_5&=&\frac{\left\langle \bar{q}q \right\rangle \left(3 m_d+7 m_u\right)}{2^{16}\times 3^2 \times 5^2 \times 7 \, \pi ^8}-\frac{m_d m_u^2 \left(3 m_d+7 m_u\right)}{2^{18}\times 3 \times 5^2 \times 7 \, \pi ^{10}}-\frac{\left\langle GG \right\rangle}{2^{21}\times 3 \times 5^2 \times 7 \, \pi ^{10}}\;,\\
c_4&=&\frac{\left\langle \bar{q}q \right\rangle m_u \left(25 m_d m_u+5 m_d^2+6 m_u^2\right)}{2^{17}\times 3^2 \times 5  \, \pi ^8}-\frac{\left\langle GG \right\rangle m_u \left(5 m_d+6 m_u\right)}{2^{22}\times 3^2 \times 5  \, \pi ^{10}}\nonumber\\
&-&\frac{\left\langle \bar{q}Gq \right\rangle \left(5 m_d+11 m_u\right)}{2^{18}\times 3^2 \times 5  \, \pi ^8}-\frac{m_d^2 m_u^4}{2^{18}\times 5  \, \pi ^{10}}-\frac{\left\langle \bar{q}q \right\rangle^2}{2^{11}\times 3^3 \times 5  \, \pi ^6} \;,\\
c_3&=& -\frac{\left\langle \bar{q}q \right\rangle^2 \left(28 m_d m_u+m_d^2+18 m_u^2\right)}{2^{12}\times 3^2 \times 5  \, \pi ^6}\nonumber\\
&+&\left\langle \bar{q}q \right\rangle \left[\frac{\left\langle GG \right\rangle \left(m_d+4 m_u\right)}{2^{16}\times 3^2 \times 5  \, \pi ^8}+\frac{m_d m_u^3 \left(11 m_d+5 m_u\right)}{2^{14}\times 3 \times 5  \, \pi ^8}+\frac{13 \left\langle \bar{q}Gq \right\rangle}{2^{13}\times 3^2 \times 5  \, \pi ^6}\right] \nonumber \\
&-&\frac{\left\langle GG \right\rangle m_d m_u^3}{2^{19}\times 3  \, \pi ^{10}}-\frac{\left\langle \bar{q}Gq \right\rangle m_u \left(19 m_d m_u+4 m_d^2+5 m_u^2\right)}{2^{15}\times 3 \times 5  \, \pi ^8}\;,\\
c_2&=&\frac{\left\langle \bar{q}q \right\rangle^3 \left(7 m_d+17 m_u\right)}{2^{10}\times 3^2 \, \pi ^4}+\left\langle \bar{q}q \right\rangle^2 \left[-\frac{m_u^2 \left(37 m_d m_u+23 m_d^2+2 m_u^2\right)}{2^{13}\times 3  \, \pi ^6}-\frac{7 \left\langle GG \right\rangle}{2^{15}\times 3^2 \, \pi ^6}\right]\nonumber\\
&+&\left\langle \bar{q}q \right\rangle \left[\frac{\left\langle GG \right\rangle m_u^2 \left(7 m_d+2 m_u\right)}{2^{17}\times 3 \, \pi ^8}+\frac{\left\langle \bar{q}Gq \right\rangle \left(52 m_d m_u+3 m_d^2+37 m_u^2\right)}{2^{14}\times 3  \, \pi ^6}\right]\nonumber\\
&-&\frac{\left\langle \bar{q}Gq \right\rangle m_d m_u^3 \left(2 m_d+m_u\right)}{2^{13} \, \pi ^8}-\frac{5 \left\langle \bar{q}Gq \right\rangle^2}{2^{15}\times 3  \, \pi ^6}-\frac{\left\langle GG \right\rangle \left\langle \bar{q}Gq \right\rangle \left(3 m_d+11 m_u\right)}{2^{19}\times 3  \, \pi ^8}\;,\\
c_1&=&\frac{\left\langle \bar{q}q \right\rangle^3 m_u \left(38 m_d m_u+9 m_d^2+9 m_u^2\right)}{2^{8}\times 3^2 \, \pi ^4}-\frac{\left\langle GG \right\rangle \left\langle \bar{q}Gq \right\rangle m_u^2 \left(3 m_d+m_u\right)}{2^{16} \, \pi ^8}\nonumber\\
&+&\left\langle \bar{q}q \right\rangle^2 \left[-\frac{11 \left\langle GG \right\rangle m_u \left(m_d+m_u\right)}{2^{13}\times 3^2 \, \pi ^6}-\frac{\left\langle \bar{q}Gq \right\rangle \left(5 m_d+12 m_u\right)}{2^{7}\times 3^2  \, \pi ^4}-\frac{m_d^2 m_u^4}{2^{8}\times 3 \, \pi ^6}\right]\nonumber\\
&+&\left\langle \bar{q}q \right\rangle \left[\frac{\left\langle \bar{q}Gq \right\rangle m_u^2 \left(32 m_d m_u+23 m_d^2+3 m_u^2\right)}{2^{11}\times 3 \, \pi ^6}+\frac{5 \left\langle GG \right\rangle \left\langle \bar{q}Gq \right\rangle}{2^{13}\times 3^2 \, \pi ^6}\right]\nonumber\\
&-&\frac{\left\langle \bar{q}Gq \right\rangle^2 \left(13 m_d m_u+m_d^2+10 m_u^2\right)}{2^{12}\times 3 \, \pi ^6}-\frac{\left\langle \bar{q}q \right\rangle^4}{2^{2}\times 3^3 \, \pi ^2}\;,\\
c_0&=&-\frac{\left\langle \bar{q}q \right\rangle^4 \left(19 m_d m_u+m_d^2+12 m_u^2\right)}{2^{6}\times 3^2 \, \pi ^2}+\frac{\left\langle \bar{q}q \right\rangle^3 m_d m_u^3 \left(3 m_d+m_u\right)}{2^{9}\, \pi ^4}\nonumber\\
&+&\left\langle \bar{q}q \right\rangle^2 \left[-\frac{\left\langle GG \right\rangle m_d m_u^3}{2^{14} \, \pi ^6}-\frac{\left\langle \bar{q}Gq \right\rangle m_u \left(94 m_d m_u+25 m_d^2+25 m_u^2\right)}{2^{10}\times 3 \, \pi ^4}\right]\nonumber\\
&+&\frac{3 \left\langle \bar{q}q \right\rangle \left\langle GG \right\rangle \left\langle \bar{q}Gq \right\rangle m_u \left(m_d+m_u\right)}{2^{14} \, \pi ^6}-\frac{\left\langle \bar{q}Gq \right\rangle^2 m_u^2 \left(8 m_d m_u+6 m_d^2+m_u^2\right)}{2^{12} \, \pi ^6}\;,
\end{eqnarray}
and
\begin{eqnarray}
\Pi^{sum}(M_B^2)=\left\langle \bar{q}q \right\rangle^4 \left(\frac{m_d m_u^3}{2^{5}\times 3^2 \, \pi ^2}+\frac{m_d^2 m_u^2}{2^{7}\, \pi ^2}-\frac{m_u^4}{2^{8}\times 3^2 \, \pi ^2}\right)+\frac{\left\langle \bar{q}q \right\rangle^2 \left\langle \bar{q}Gq \right\rangle \left(m_d^2 m_u^3\right)}{2^{8}\times 3 \, \pi ^4}\;.
\end{eqnarray}

\subsection{The spectral densities of $1^{++}$ $p$-$\bar{p}$ hexaquark states}
The $1^{++}$ $p$-$\bar{p}$ hexaquark state spectral densities on the OPE side can be expressed as:
\begin{eqnarray}
c_7&=&-\frac{1}{2^{20}\times 3^4\times 5^2 \times 7^2 \, \pi ^{10}}\;,\\
c_6&=&-\frac{  m_u \left(7 m_d+4 m_u\right)}{2^{22}\times 3^2\times 5^2 \times 7 \, \pi ^{10}}\;,\\
c_5&=&\frac{\left\langle \bar{q}q \right\rangle \left(3 m_d+7 m_u\right)}{2^{16}\times 3^2\times 5^2 \times 7 \, \pi ^8}-\frac{m_d m_u^2 \left(3 m_d+7 m_u\right)}{2^{18}\times 3 \times 5^2 \times 7 \, \pi ^{10}}-\frac{\left\langle GG \right\rangle^{2}}{2^{21}\times 3 \times 5^2 \times 7 \, \pi ^{10}}\;,\\
c_4&=&\frac{\left\langle \bar{q}q \right\rangle m_u \left(25 m_d m_u+5 m_d^2+6 m_u^2\right)}{2^{17}\times 3^2 \times 5  \, \pi ^8}-\frac{\left\langle GG \right\rangle^{2} m_u \left(5 m_d+6 m_u\right)}{2^{22}\times 3^2 \times 5  \, \pi ^{10}}\nonumber\\
&-&\frac{\left\langle \bar{q}Gq \right\rangle \left(5 m_d+11 m_u\right)}{2^{18}\times 3^2\times 5\, \pi ^8}-\frac{27 m_d^2 m_u^4+2^{7} \, \pi ^4 \left\langle \bar{q}q \right\rangle^{2}}{2^{18}\times 3^3 \times 5   \, \pi ^{10}}\;,\\
c_3&=&\left\langle \bar{q}q \right\rangle \left[\frac{\left\langle GG \right\rangle^{2} \left(m_d+4 m_u\right)}{2^{16}\times 3^2 \times 5  \, \pi ^8}+\frac{m_d m_u^3 \left(11 m_d+5 m_u\right)}{2^{14}\times 3 \times 5   \, \pi ^8}+\frac{13 \left\langle \bar{q}Gq \right\rangle}{2^{13}\times 3^2 \times 5  \, \pi ^6}\right]-\frac{\left\langle GG \right\rangle^{2} m_d m_u^3}{2^{19}\times 3 \, \pi ^{10}}\nonumber\\
&-&\frac{\left\langle \bar{q}Gq \right\rangle m_u \left(19 m_d m_u+4 m_d^2+5 m_u^2\right)}{2^{15}\times 3 \times 5  \, \pi ^8}-\frac{\left\langle \bar{q}q \right\rangle^{2} \left(28 m_d m_u+m_d^2+18 m_u^2\right)}{2^{12}\times 3^2 \times 5  \, \pi ^6}\;,\\
c_2&=&\frac{\left\langle \bar{q}q \right\rangle^3 \left(7 m_d+17 m_u\right)}{2^{10}\times 3^2   \, \pi ^4}-\frac{7 \left\langle \bar{q}q \right\rangle^2 \left\langle GG \right\rangle^{2}}{2^{15}\times 3^2   \, \pi ^6}-\frac{5 \left\langle \bar{q}Gq \right\rangle^2}{2^{15}\times 3 \, \pi ^6}-\frac{\left\langle \bar{q}Gq \right\rangle m_d m_u^3 \left(2 m_d+m_u\right)}{2^{13}\, \pi ^8}\nonumber\\
&+&\left\langle \bar{q}q \right\rangle \left[\left\langle GG \right\rangle^{2} \left(\frac{m_u^2 \left(7 m_d+2 m_u\right)}{2^{17}\times 3 \, \pi ^8}-\frac{7 \left\langle \bar{q}Gq \right\rangle}{2^{15}\times 3^2   \, \pi ^6}\right)+\frac{\left\langle \bar{q}Gq \right\rangle \left(52 m_d m_u+3 m_d^2+37 m_u^2\right)}{2^{14}\times 3 \, \pi ^6}\right]\nonumber\\
&-&\frac{\left\langle GG \right\rangle^{2} \left\langle \bar{q}Gq \right\rangle \left(3 m_d+11 m_u\right)}{2^{19}\times 3 \, \pi ^8}-\frac{\left\langle \bar{q}q \right\rangle^{2} m_u^2 \left(37 m_d m_u+23 m_d^2+2 m_u^2\right)}{2^{13}\times 3   \, \pi ^6}\;,\\
c_1&=&\frac{\left\langle \bar{q}q \right\rangle^3 m_u \left(38 m_d m_u+9 m_d^2+9 m_u^2\right)}{2^{8}\times 3^2 \, \pi ^4}+\left\langle \bar{q}q \right\rangle^2 \big(-\frac{11 \left\langle GG \right\rangle^{2} m_u \left(m_d+m_u\right)}{2^{13}\times 3^2   \, \pi ^6}\nonumber\\
&-&\frac{\left\langle \bar{q}Gq \right\rangle \left(5 m_d+12 m_u\right)}{2^{7}\times 3^2   \, \pi ^4}\big)+\left\langle \bar{q}q \right\rangle \bigg[\frac{\left\langle \bar{q}Gq \right\rangle m_u^2 \left(32 m_d m_u+23 m_d^2+3 m_u^2\right)}{2^{11}\times 3  \, \pi ^6}\nonumber\\
&-&\frac{11 \left\langle GG \right\rangle^{2} \left\langle \bar{q}Gq \right\rangle m_u \left(m_d+m_u\right)}{2^{13}\times 3^2   \, \pi ^6}\bigg]-\frac{\left\langle \bar{q}q \right\rangle^4}{2^{2}\times 3^3 \, \pi ^2}\nonumber\\
&-&\frac{\left\langle GG \right\rangle^{2} \left\langle \bar{q}Gq \right\rangle m_u^2 \left(3 m_d+m_u\right)}{2^{16}\, \pi ^8}-\frac{\left\langle \bar{q}Gq \right\rangle^2 \left(13 m_d m_u+m_d^2+10 m_u^2\right)}{2^{12}\times 3 \, \pi ^6}-\frac{\left\langle \bar{q}q \right\rangle^{2} m_d^2 m_u^4}{2^{8}\times 3 \, \pi ^6}\;,\\
c_0&=&-\frac{\left\langle \bar{q}q \right\rangle^4 \left(19 m_d m_u+m_d^2+12 m_u^2\right)}{2^{6}\times 3^2 \, \pi ^2}+\frac{\left\langle \bar{q}q \right\rangle^3 m_d m_u^3 \left(3 m_d+m_u\right)}{2^{9} \, \pi ^4}\nonumber\\
&+&\left\langle \bar{q}q \right\rangle^2 \left[-\frac{\left\langle GG \right\rangle^{2} m_d m_u^3}{2^{14}\, \pi ^6}-\frac{\left\langle \bar{q}Gq \right\rangle m_u \left(94 m_d m_u+25 m_d^2+25 m_u^2\right)}{2^{10}\times 3 \, \pi ^4}\right]\nonumber\\
&-&\frac{\left\langle \bar{q}q \right\rangle \left\langle GG \right\rangle^{2} \left\langle \bar{q}Gq \right\rangle m_d m_u^3}{2^{14}\, \pi ^6}-\frac{\left\langle \bar{q}Gq \right\rangle^2 m_u^2 \left(8 m_d m_u+6 m_d^2+m_u^2\right)}{2^{12} \, \pi ^6}\;,
\end{eqnarray}
and
\begin{eqnarray}
\Pi^{sum}(M_B^2)&=&\left\langle \bar{q}q \right\rangle^{2} \left\langle \bar{q}Gq \right\rangle \left(-\frac{m_d m_u^4}{2^{9}\times 3 \, \pi ^4}-\frac{m_d^2 m_u^3}{2^{9} \, \pi ^4}\right)\nonumber\\
&+&\frac{\left\langle \bar{q}q \right\rangle^4 \left(m_d m_u^2 \left(m_d + m_u\right)\right)}{2^{7}\times 3^2   \, \pi ^2}\;.
\end{eqnarray}

\subsection{The spectral densities of $0^{-+}$ $\Lambda$-$\bar{\Lambda}$ hexaquark states}
The $0^{-+}$ $\Lambda$-$\bar{\Lambda}$ hexaquark state spectral densities on the OPE side can be expressed as:
\begin{eqnarray}
c_7&=&\frac{1}{2^{23}\times 3^2 \times 5^2 \times 7^2 \, \pi ^{10}}\;,\\
c_6&=&\frac{ m_s^2 }{2^{20}\times 3^2 \times 5^2 \times 7 \, \pi ^{10}}\;,\\
c_5&=&\frac{\left\langle GG \right\rangle}{2^{22}\times 3^2 \times 5^2  \, \pi ^{10}}-\frac{\left\langle \bar{s}s \right\rangle m_s}{2^{17}\times 3^2 \times 5^2  \, \pi ^8}\;,\\
c_4&=&\frac{\left\langle GG \right\rangle m_s^2}{2^{21}\times 3 \times 5  \, \pi ^{10}}+\frac{\left\langle \bar{s}Gs \right\rangle m_s}{2^{17}\times 3 \times 5  \, \pi ^8}+\frac{\left\langle \bar{q}q \right\rangle^2}{2^{13}\times 3^2 \times 5  \, \pi ^6}+\frac{\left\langle \bar{s}s \right\rangle^2}{2^{14}\times 3^2 \times 5\, \pi ^6} \;,\\
c_3&=&\frac{\left\langle \bar{q}q \right\rangle^2 m_s^2}{2^{11}\times 3^2 \, \pi ^6}+\frac{\left\langle \bar{s}s \right\rangle^2 m_s^2}{2^{14}\times 3^2  \, \pi ^6}-\frac{\left\langle \bar{s}s \right\rangle \left\langle GG \right\rangle m_s}{2^{17}\times 3^2 \, \pi ^8}-\frac{\left\langle \bar{q}q \right\rangle \left\langle \bar{q}Gq \right\rangle}{2^{12}\times 3^2 \, \pi ^6}-\frac{\left\langle \bar{s}s \right\rangle \left\langle \bar{s}Gs \right\rangle}{2^{13}\times 3^2 \, \pi ^6} \;,\\
c_2&=&\left\langle \bar{q}q \right\rangle^2 \left[\frac{\left\langle GG \right\rangle}{2^{13}\times 3^2 \, \pi ^6}-\frac{\left\langle \bar{s}s \right\rangle m_s}{2^{7}\times 3^2  \, \pi ^4}\right]-\frac{\left\langle \bar{q}q \right\rangle \left\langle \bar{q}Gq \right\rangle m_s^2}{2^{10}\times 3 \, \pi ^6}\nonumber\\
&+&\left\langle GG \right\rangle \left[\frac{\left\langle \bar{s}Gs \right\rangle m_s}{2^{16}\times 3 \, \pi ^8}+\frac{\left\langle \bar{s}s \right\rangle^2}{2^{13}\times 3^2 \, \pi ^6}\right]+\frac{\left\langle \bar{q}Gq \right\rangle^2}{2^{13}\times 3  \, \pi ^6}+\frac{\left\langle \bar{s}Gs \right\rangle^2}{2^{14}\times 3  \, \pi ^6}\;,\\
c_1&=&\left\langle \bar{q}q \right\rangle^2 \left[\frac{\left\langle GG \right\rangle m_s^2}{2^{12}\times 3 \, \pi ^6}+\frac{\left\langle \bar{s}Gs \right\rangle m_s}{2^{7}\times 3 \, \pi ^4}+\frac{\left\langle \bar{s}s \right\rangle^2}{2^{4}\times 3^2 \, \pi ^2}\right]+\left\langle \bar{q}q \right\rangle \left[\frac{\left\langle \bar{s}s \right\rangle \left\langle \bar{q}Gq \right\rangle m_s}{2^{9}\, \pi ^4}-\frac{\left\langle GG \right\rangle \left\langle \bar{q}Gq \right\rangle}{2^{13}\times 3  \, \pi ^6}\right]\nonumber\\
&+&\left\langle GG \right\rangle \left[\frac{\left\langle \bar{s}s \right\rangle^2 m_s^2}{2^{14}\times 3  \, \pi ^6}-\frac{\left\langle \bar{s}s \right\rangle \left\langle \bar{s}Gs \right\rangle}{2^{13}\times 3  \, \pi ^6}\right]+\frac{\left\langle \bar{q}Gq \right\rangle^2 m_s^2}{2^{12} \, \pi ^6}+\frac{\left\langle \bar{q}q \right\rangle^4}{2^{5}\times 3^2 \, \pi ^2}\;,\\
c_0&=&\frac{\left\langle \bar{q}q \right\rangle^4 m_s^2}{2^{4}\times 3^2 \, \pi ^2}+\frac{\left\langle \bar{q}q \right\rangle^2 \left\langle \bar{s}s \right\rangle^2 m_s^2}{2^{5}\times 3^2 \, \pi ^2}-\frac{\left\langle \bar{q}q \right\rangle \left\langle GG \right\rangle \left\langle \bar{q}Gq \right\rangle m_s^2}{2^{12}\times 3 \, \pi ^6}\;.
\end{eqnarray}

It should be noted that, the contribution of the masses of $u$ and $d$ quarks are so tiny for $\Lambda$-$\bar{\Lambda}$, $\Sigma$-$\bar{\Sigma}$ and $\Xi$-$\bar{\Xi}$ states that we have not displayed them.

\subsection{The spectral densities of $0^{++}$ $\Lambda$-$\bar{\Lambda}$ hexaquark states}
The $0^{++}$ $\Lambda$-$\bar{\Lambda}$ hexaquark state spectral densities on the OPE side can be expressed as:
\begin{eqnarray}
c_7&=&-\frac{1}{2^{23}\times 3^2\times 5^2\times 7^2\, \pi ^{10}}\;,\\
c_6&=&\frac{ m_s^2}{2^{20}\times 3^2\times 5^2\times 7^1\, \pi ^{10}}\;,\\
c_5&=&-\frac{\left\langle \bar{s}s \right\rangle m_s}{2^{17}\times 3 \times 5^2\, \pi ^8}-\frac{\left\langle GG \right\rangle^{2}}{2^{22}\times 3^2\times 5^2\, \pi ^{10}}\;,\\
c_4&=&\frac{\left\langle GG \right\rangle^{2} m_s^2}{2^{21}\times 3\times 5\, \pi ^{10}}+\frac{3 \left\langle \bar{s}Gs \right\rangle m_s+8 \pi ^2 (\left\langle \bar{s}s \right\rangle^{2}-2 \left\langle \bar{q}q \right\rangle^{2})}{2^{17}\times 3^2 \times 5  \, \pi ^8}\;,\\
c_3&=&-\frac{\left\langle \bar{s}s \right\rangle \left\langle GG \right\rangle^{2} m_s}{2^{17}\times 3 \, \pi ^8}-\frac{m_s^2 (\left\langle \bar{s}s \right\rangle^{2}-8 \left\langle \bar{q}q \right\rangle^{2})+2 \left\langle \bar{s}s \right\rangle \left\langle \bar{s}Gs \right\rangle}{2^{14}\times 3^2 \, \pi ^6}+\frac{\left\langle \bar{q}q \right\rangle \left\langle \bar{q}Gq \right\rangle}{2^{12}\times 3^2 \, \pi ^6}\;,\\
c_2&=&\frac{\left\langle GG \right\rangle^{2} \left(3 \left\langle \bar{s}Gs \right\rangle m_s+8 \pi ^2 \left\langle \bar{s}s \right\rangle^2\right)}{2^{16}\times 3^2 \, \pi ^8}+\frac{\left\langle \bar{s}Gs \right\rangle^2-2^{7} \, \pi ^2 \left\langle \bar{s}s \right\rangle \left\langle \bar{q}q \right\rangle^{2} m_s}{2^{14}\times 3 \, \pi ^6}\nonumber\\
&-&\frac{\left\langle \bar{q}q \right\rangle^2 \left\langle GG \right\rangle^{2}}{2^{13}\times 3^2   \, \pi ^6}-\frac{\left\langle \bar{q}Gq \right\rangle^2}{2^{13}\times 3   \, \pi ^6}-\frac{\left\langle \bar{q}q \right\rangle \left\langle \bar{q}Gq \right\rangle m_s^2}{2^{10}\times 3 \, \pi ^6}\;,\\
c_1&=&\left\langle \bar{q}q \right\rangle^2 \left[\frac{\left\langle GG \right\rangle^{2} m_s^2}{2^{12}\times 3 \, \pi ^6}+\frac{\left\langle \bar{s}s \right\rangle^2}{2^{4}\times 3^2 \, \pi ^2}\right]+\left\langle \bar{q}q \right\rangle \left[\frac{\left\langle \bar{s}s \right\rangle \left\langle \bar{q}Gq \right\rangle m_s}{2^{7} \, \pi ^4}+\frac{\left\langle GG \right\rangle^{2} \left\langle \bar{q}Gq \right\rangle}{2^{13}\times 3   \, \pi ^6}\right]\nonumber\\
&-&\frac{\left\langle \bar{s}s \right\rangle \left\langle GG \right\rangle^{2} \left(\left\langle \bar{s}s \right\rangle m_s^2+2 \left\langle \bar{s}Gs \right\rangle\right)}{2^{14}\times 3 \, \pi ^6}+\frac{\left\langle \bar{q}Gq \right\rangle^2 m_s^2}{2^{12} \, \pi ^6}+\frac{\left\langle \bar{s}Gs \right\rangle \left\langle \bar{q}q \right\rangle^{2} m_s}{2^{7}\times 3 \, \pi ^4}-\frac{\left\langle \bar{q}q \right\rangle^4}{2^{5}\times 3^2\, \pi ^2}\;,\\
c_0&=&\frac{\left\langle \bar{q}q \right\rangle^4 m_s^2}{2^{4}\times 3^2 \, \pi ^2}-\frac{\left\langle \bar{q}q \right\rangle^2 \left\langle \bar{s}s \right\rangle^2 m_s^2}{2^{5}\times 3^2\, \pi ^2}-\frac{\left\langle \bar{q}q \right\rangle \left\langle GG \right\rangle^{2} \left\langle \bar{q}Gq \right\rangle m_s^2}{2^{12}\times 3 \, \pi ^6}\;,
\end{eqnarray}

\subsection{The spectral densities of $1^{--}$ $\Lambda$-$\bar{\Lambda}$ hexaquark states}
The $1^{--}$ $\Lambda$-$\bar{\Lambda}$ hexaquark state spectral densities on the OPE side can be expressed as:
\begin{eqnarray}
c_7&=&-\frac{1}{2^{20}\times 3^4\times 5^2 \times 7^2 \, \pi ^{10}}\;,\\
c_6&=&-\frac{m_s^2  }{2^{20}\times 3^2 \times 5^2 \times 7 \, \pi ^{10}}\;,\\
c_5&=&\frac{\left\langle \bar{s}s \right\rangle m_s}{2^{14}\times 3^2 \times 5^2 \times 7 \, \pi ^8}-\frac{\left\langle GG \right\rangle}{2^{21}\times 3 \times 5^2 \times 7 \, \pi ^{10}}\;,\\
c_4&=&-\frac{\left\langle GG \right\rangle m_s^2}{2^{21}\times 3 \times 5  \, \pi ^{10}}-\frac{\left\langle \bar{s}Gs \right\rangle m_s}{2^{17}\times 3 \times 5  \, \pi ^8}-\frac{\left\langle \bar{q}q \right\rangle^2}{2^{14}\times 3^3  \, \pi ^6}-\frac{\left\langle \bar{s}s \right\rangle^2}{2^{14}\times 3^2 \times 5\, \pi ^6}\;,\\
c_3&=&-\frac{\left\langle \bar{q}q \right\rangle^2 m_s^2}{2^{11}\times 3^2 \, \pi ^6}-\frac{\left\langle \bar{s}s \right\rangle^2 m_s^2}{2^{12}\times 3^2 \times 5  \, \pi ^6}+\frac{\left\langle \bar{s}s \right\rangle \left\langle GG \right\rangle m_s}{2^{16}\times 3 \times 5  \, \pi ^8}\nonumber\\
&+&\frac{\left\langle \bar{q}q \right\rangle \left\langle \bar{q}Gq \right\rangle}{2^{10}\times 3^2 \times 5 \, \pi ^6}+\frac{\left\langle \bar{s}s \right\rangle \left\langle \bar{s}Gs \right\rangle}{2^{13}\times 3^2 \, \pi ^6}\;,\\
c_2&=&\left\langle \bar{q}q \right\rangle^2 \left[\frac{5 \left\langle \bar{s}s \right\rangle m_s}{2^{9}\times 3^2  \, \pi ^4}-\frac{\left\langle GG \right\rangle}{2^{15}\times 3  \, \pi ^6}\right]+\frac{\left\langle \bar{q}q \right\rangle \left\langle \bar{q}Gq \right\rangle m_s^2}{2^{10}\times 3 \, \pi ^6}\nonumber\\
&+&\left\langle GG \right\rangle \left[-\frac{\left\langle \bar{s}Gs \right\rangle m_s}{2^{16}\times 3 \, \pi ^8}-\frac{\left\langle \bar{s}s \right\rangle^2}{2^{13}\times 3^2 \, \pi ^6}\right]-\frac{\left\langle \bar{q}Gq \right\rangle^2}{2^{15}\, \pi ^6}-\frac{\left\langle \bar{s}Gs \right\rangle^2}{2^{14}\times 3  \, \pi ^6}  \;,\\
c_1&=&\left\langle \bar{q}q \right\rangle^2 \left[-\frac{\left\langle GG \right\rangle m_s^2}{2^{12}\times 3 \, \pi ^6}-\frac{\left\langle \bar{q}Gq \right\rangle m_s}{2^{7}\times 3 \, \pi ^4}-\frac{\left\langle \bar{s}Gs \right\rangle m_s}{2^{7}\times 3 \, \pi ^4}-\frac{\left\langle \bar{s}s \right\rangle^2}{2^{4}\times 3^2 \, \pi ^2}\right]\nonumber\\
&+&\left\langle \bar{q}q \right\rangle \left[\frac{\left\langle GG \right\rangle \left\langle \bar{q}Gq \right\rangle}{2^{12}\times 3^2 \, \pi ^6}-\frac{\left\langle \bar{s}s \right\rangle \left\langle \bar{q}Gq \right\rangle m_s}{2^{5}\times 3^2 \, \pi ^4}\right]\nonumber\\
&+&\left\langle GG \right\rangle \left[\frac{\left\langle \bar{s}s \right\rangle \left\langle \bar{s}Gs \right\rangle}{2^{13}\times 3  \, \pi ^6}-\frac{\left\langle \bar{s}s \right\rangle^2 m_s^2}{2^{13}\times 3^2 \, \pi ^6}\right]-\frac{\left\langle \bar{q}Gq \right\rangle^2 m_s^2}{2^{12} \, \pi ^6}-\frac{\left\langle \bar{q}q \right\rangle^4}{2^{4}\times 3^3  \, \pi ^2}\;,\\
c_0&=&-\frac{\left\langle \bar{q}q \right\rangle^4 m_s^2}{2^{4}\times 3^2 \, \pi ^2}-\frac{\left\langle \bar{q}q \right\rangle^2 \left\langle \bar{s}s \right\rangle^2 m_s^2}{2^{6}\times 3^2 \, \pi ^2}+\frac{\left\langle \bar{q}q \right\rangle \left\langle GG \right\rangle \left\langle \bar{q}Gq \right\rangle m_s^2}{2^{12}\times 3 \, \pi ^6}\;,
\end{eqnarray}

\subsection{The spectral densities of $1^{++}$ $\Lambda$-$\bar{\Lambda}$ hexaquark states}
The $1^{++}$ $\Lambda$-$\bar{\Lambda}$ hexaquark state spectral densities on the OPE side can be expressed as:
\begin{eqnarray}
c_7&=&-\frac{1}{2^{20}\times 3^4\times 5^2 \times 7^2 \, \pi ^{10}}\;,\\
c_6&=&\frac{  m_s^2}{2^{20}\times 3^2\times 5^2\times 7^1\, \pi ^{10}}\;,\\
c_5&=&-\frac{\left\langle \bar{s}s \right\rangle m_s}{2^{15}\times 3^2 \times 5 \times 7  \, \pi ^8}-\frac{\left\langle GG \right\rangle^{2}}{2^{21}\times 3 \times 5^2 \times 7 \, \pi ^{10}}\;,\\
c_4&=&\frac{\left\langle GG \right\rangle^{2} m_s^2}{2^{21}\times 3\times 5\, \pi ^{10}}+\frac{9 \left\langle \bar{s}Gs \right\rangle m_s+8 \pi ^2 (3 \left\langle \bar{s}s \right\rangle^{2}-5 \left\langle \bar{q}q \right\rangle^{2})}{2^{17}\times 3^3 \times 5  \, \pi ^8}\;,\\
c_3&=&-\frac{7 \left\langle \bar{s}s \right\rangle \left\langle GG \right\rangle^{2} m_s}{2^{16}\times 3^2 \times 5  \, \pi ^8}-\frac{-20 \left\langle \bar{q}q \right\rangle^{2} m_s^2+2 \left\langle \bar{s}s \right\rangle^{2} m_s^2+5 \left\langle \bar{s}s \right\rangle \left\langle \bar{s}Gs \right\rangle}{2^{13}\times 3^2 \times 5  \, \pi ^6}+\frac{\left\langle \bar{q}q \right\rangle \left\langle \bar{q}Gq \right\rangle}{2^{10}\times 3^2 \times 5  \, \pi ^6}\;,\\
c_2&=&-\frac{\left\langle \bar{q}q \right\rangle \left\langle \bar{q}Gq \right\rangle m_s^2}{2^{10}\times 3 \, \pi ^6}+\frac{3 \left\langle \bar{s}Gs \right\rangle^2-352 \pi ^2 \left\langle \bar{s}s \right\rangle \left\langle \bar{q}q \right\rangle^{2} m_s}{2^{14}\times 3^2 \, \pi ^6}\nonumber\\
&+&\frac{\left\langle GG \right\rangle^{2} \left[3 \left\langle \bar{s}Gs \right\rangle m_s-6 \pi ^2 \left\langle \bar{q}q \right\rangle^{2}+8 \pi ^2 \left\langle \bar{s}s \right\rangle^{2}\right]}{2^{16}\times 3^2 \, \pi ^8}-\frac{\left\langle \bar{q}Gq \right\rangle^2}{2^{15} \, \pi ^6}\;,\\
c_1&=&\left\langle \bar{q}q \right\rangle \left[\frac{\left\langle \bar{s}s \right\rangle \left\langle \bar{q}Gq \right\rangle m_s}{2^{4}\times 3^2 \, \pi ^4}+\frac{\left\langle GG \right\rangle^{2} \left\langle \bar{q}Gq \right\rangle}{2^{12}\times 3^2 \, \pi ^6}\right]+\frac{\left\langle GG \right\rangle^{2} \left[m_s^2 (6 \left\langle \bar{q}q \right\rangle^{2}-\left\langle \bar{s}s \right\rangle^{2})+3 \left\langle \bar{s}s \right\rangle \left\langle \bar{s}Gs \right\rangle\right]}{2^{13}\times 3^2   \, \pi ^6}\nonumber\\
&+&\frac{\left\langle \bar{q}Gq \right\rangle^2 m_s^2}{2^{12} \, \pi ^6}+\frac{\left\langle \bar{s}Gs \right\rangle \left\langle \bar{q}q \right\rangle^{2} m_s}{2^{7}\times 3 \, \pi ^4}-\frac{\left\langle \bar{q}q \right\rangle^4}{2^{4}\times 3^3 \, \pi ^2}-\frac{\left\langle \bar{q}q \right\rangle^2 \left\langle \bar{s}s \right\rangle^2}{2^{4}\times 3^2 \, \pi ^2}\;,\\
c_0&=&-\frac{\left\langle \bar{q}q \right\rangle^4 m_s^2}{2^{4}\times 3^2 \, \pi ^2}-\frac{\left\langle \bar{q}q \right\rangle^2 \left\langle \bar{s}s \right\rangle^2 m_s^2}{2^{6}\times 3^2 \, \pi ^2}+\frac{\left\langle \bar{q}q \right\rangle \left\langle GG \right\rangle^{2} \left\langle \bar{q}Gq \right\rangle m_s^2}{2^{12}\times 3 \, \pi ^6}\;,
\end{eqnarray}

\subsection{The spectral densities of $0^{-+}$ $\Sigma$-$\bar{\Sigma}$ hexaquark states}
The $0^{-+}$ $\Sigma$-$\bar{\Sigma}$ hexaquark state spectral densities on the OPE side can be expressed as:
\begin{eqnarray}
c_7&=&\frac{1}{2^{23}\times 3^2 \times 5^2 \times 7^2 \, \pi ^{10}}\;,\\
c_6&=&0\;,\\
c_5&=&-\frac{\left\langle \bar{q}q \right\rangle m_s}{2^{16} \times 3^2 \times 5^2 \, \pi ^8}+\frac{\left\langle \bar{s}s \right\rangle m_s}{2^{17}\times 3^2 \times 5^2  \, \pi ^8}+\frac{\left\langle GG \right\rangle}{2^{22}\times 3^2 \times 5^2  \, \pi ^{10}} \;,\\
c_4&=&\frac{\left\langle \bar{q}Gq \right\rangle m_s}{2^{17}\times 3 \times 5  \, \pi ^8}+\frac{\left\langle \bar{q}q \right\rangle^2}{2^{14}\times 3^2 \times 5\, \pi ^6}+\frac{\left\langle \bar{q}q \right\rangle \left\langle \bar{s}s \right\rangle}{2^{13}\times 3^2 \times 5  \, \pi ^6}\;,\\
c_3&=&\frac{\left\langle \bar{q}q \right\rangle^2 m_s^2}{2^{12}\times 3^2 \, \pi ^6}+\left\langle \bar{q}q \right\rangle \left[-\frac{\left\langle \bar{s}s \right\rangle m_s^2}{2^{12}\times 3^2 \, \pi ^6}-\frac{\left\langle GG \right\rangle m_s}{2^{17}\times 3^2 \, \pi ^8}-\frac{\left\langle \bar{q}Gq \right\rangle}{2^{13}\times 3^2 \, \pi ^6}-\frac{\left\langle \bar{s}Gs \right\rangle}{2^{13}\times 3^2 \, \pi ^6}\right]\nonumber\\
&+&\frac{\left\langle \bar{s}s \right\rangle^2 m_s^2}{2^{14}\times 3^2  \, \pi ^6}+\frac{\left\langle \bar{s}s \right\rangle \left\langle GG \right\rangle m_s}{2^{18}\times 3^2  \, \pi ^8}-\frac{\left\langle \bar{s}s \right\rangle \left\langle \bar{q}Gq \right\rangle}{2^{13}\times 3^2 \, \pi ^6}\;,\\
c_2&=&\left\langle \bar{q}q \right\rangle \left[\frac{\left\langle \bar{s}s \right\rangle^2 m_s}{2^{8}\times 3^2 \, \pi ^4}-\frac{\left\langle \bar{q}Gq \right\rangle m_s^2}{2^{11}\times 3 \, \pi ^6}+\frac{\left\langle \bar{s}s \right\rangle \left\langle GG \right\rangle}{2^{13}\times 3^2 \, \pi ^6}\right]-\frac{\left\langle \bar{q}q \right\rangle^3 m_s}{2^{7}\times 3^2  \, \pi ^4}\nonumber\\
&+&\left\langle \bar{q}Gq \right\rangle \left[\frac{\left\langle \bar{s}s \right\rangle m_s^2}{2^{12}\times 3 \, \pi ^6}+\frac{\left\langle \bar{s}Gs \right\rangle}{2^{13}\times 3  \, \pi ^6}\right]+\left\langle \bar{q}q \right\rangle^2 \left[\frac{\left\langle GG \right\rangle}{2^{13}\times 3^2 \, \pi ^6}-\frac{\left\langle \bar{s}s \right\rangle m_s}{2^{8}\times 3^2 \, \pi ^4}\right]\nonumber\\
&+&\frac{\left\langle GG \right\rangle \left\langle \bar{q}Gq \right\rangle m_s}{2^{17}\times 3 \, \pi ^8}+\frac{\left\langle \bar{q}Gq \right\rangle^2}{2^{14}\times 3  \, \pi ^6}\;,\\
c_1&=&\left\langle \bar{q}q \right\rangle \left[\frac{\left\langle \bar{s}s \right\rangle \left\langle \bar{q}Gq \right\rangle m_s}{2^{8}\times 3 \, \pi ^4}-\frac{\left\langle \bar{s}s \right\rangle \left\langle \bar{s}Gs \right\rangle m_s}{2^{9}\times 3 \, \pi ^4}-\left\langle GG \right\rangle \big(\frac{\left\langle \bar{q}Gq \right\rangle}{2^{13}\times 3  \, \pi ^6}+\frac{\left\langle \bar{s}Gs \right\rangle}{2^{14}\times 3  \, \pi ^6}\big)\right]\nonumber\\
&+&\left\langle \bar{q}q \right\rangle^2 \left[\frac{\left\langle \bar{q}Gq \right\rangle m_s}{2^{8} \, \pi ^4}+\frac{\left\langle \bar{s}Gs \right\rangle m_s}{2^{8}\times 3 \, \pi ^4}+\frac{\left\langle \bar{s}s \right\rangle^2}{2^{5}\times 3^2 \, \pi ^2}\right]+\frac{\left\langle \bar{q}Gq \right\rangle^2 m_s^2}{2^{13} \, \pi ^6}\nonumber\\
&-&\frac{\left\langle \bar{s}s \right\rangle^2 \left\langle \bar{q}Gq \right\rangle m_s}{2^{9}\times 3 \, \pi ^4}+\frac{\left\langle \bar{q}q \right\rangle^3 \left\langle \bar{s}s \right\rangle}{2^{4}\times 3^2 \, \pi ^2}-\frac{\left\langle \bar{s}s \right\rangle \left\langle GG \right\rangle \left\langle \bar{q}Gq \right\rangle}{2^{14}\times 3  \, \pi ^6}\;,\\
c_0&=&\frac{\left\langle \bar{q}q \right\rangle^4 m_s^2}{2^{4}\times 3^2 \, \pi ^2}-\frac{\left\langle \bar{q}q \right\rangle^3 \left\langle \bar{s}s \right\rangle m_s^2}{2^{4}\times 3^2 \, \pi ^2}+\frac{\left\langle \bar{q}q \right\rangle^2 \left\langle \bar{s}s \right\rangle^2 m_s^2}{2^{6}\times 3^2 \, \pi ^2}\;,
\end{eqnarray}

\subsection{The spectral densities of $0^{++}$ $\Sigma$-$\bar{\Sigma}$ hexaquark states}
The $0^{++}$ $\Sigma$-$\bar{\Sigma}$ hexaquark state spectral densities on the OPE side can be expressed as:
\begin{eqnarray}
c_7&=&-\frac{1}{2^{23}\times 3^2\times 5^2\times 7^2\, \pi ^{10}}\;,\\
c_6&=&0\;,\\
c_5&=&\frac{\left\langle \bar{q}q \right\rangle m_s}{2^{16}\times 3^2 \times 5^2 \, \pi ^8}-\frac{\left\langle \bar{s}s \right\rangle m_s}{2^{17}\times 3^2\times 5^2\, \pi ^8}-\frac{\left\langle GG \right\rangle^{2}}{2^{22}\times 3^2\times 5^2\, \pi ^{10}}\;,\\
c_4&=&-\frac{\left\langle \bar{q}Gq \right\rangle m_s}{2^{17}\times 3\times 5\, \pi ^8}-\frac{\left\langle \bar{q}q \right\rangle \left\langle \bar{s}s \right\rangle}{2^{13}\times 3^2 \times 5  \, \pi ^6}+\frac{\left\langle \bar{q}q \right\rangle^{2}}{2^{14}\times 3^2 \times 5 \, \pi ^6}\;,\\
c_3&=&\left\langle \bar{q}q \right\rangle \left[\frac{2 \left\langle \bar{s}s \right\rangle m_s^2+\left\langle \bar{s}Gs \right\rangle}{2^{13}\times 3^2   \, \pi ^6}+\frac{\left\langle GG \right\rangle^{2} m_s}{2^{17}\times 3^2  \, \pi ^8}-\frac{\left\langle \bar{q}Gq \right\rangle}{2^{13}\times 3^2   \, \pi ^6}\right]-\frac{\left\langle \bar{s}s \right\rangle \left\langle GG \right\rangle^{2} m_s}{2^{18}\times 3^2\, \pi ^8}\nonumber\\
&-&\frac{m_s^2 (4 \left\langle \bar{q}q \right\rangle^{2}+\left\langle \bar{s}s \right\rangle^{2})}{2^{14}\times 3^2 \, \pi ^6}+\frac{\left\langle \bar{s}s \right\rangle \left\langle \bar{q}Gq \right\rangle}{2^{13}\times 3^2   \, \pi ^6}\;,\\
c_2&=&-\frac{\left\langle \bar{q}q \right\rangle^3 m_s}{2^{7}\times 3^2   \, \pi ^4}+\left\langle \bar{q}q \right\rangle \left[\frac{\left\langle \bar{q}Gq \right\rangle m_s^2}{2^{11}\times 3  \, \pi ^6}-\frac{\left\langle \bar{s}s \right\rangle^{2} m_s}{2^{8}\times 3^2 \, \pi ^4}-\frac{\left\langle \bar{s}s \right\rangle \left\langle GG \right\rangle^{2}}{2^{13}\times 3^2   \, \pi ^6}\right]+\frac{\left\langle \bar{s}s \right\rangle \left\langle \bar{q}q \right\rangle^{2} m_s}{2^{8}\times 3 \, \pi ^4}\nonumber\\
&-&\frac{\left\langle \bar{q}Gq \right\rangle \left(2 \left\langle \bar{s}s \right\rangle m_s^2+\left\langle \bar{s}Gs \right\rangle\right)}{2^{13}\times 3   \, \pi ^6}-\frac{\left\langle GG \right\rangle^{2} \left\langle \bar{q}Gq \right\rangle m_s}{2^{17}\times 3 \, \pi ^8}+\frac{\left\langle \bar{q}q \right\rangle^2 \left\langle GG \right\rangle^{2}}{2^{13}\times 3^2   \, \pi ^6}+\frac{\left\langle \bar{q}Gq \right\rangle^2}{2^{14}\times 3 \, \pi ^6}\;,\\
c_1&=&\left\langle \bar{q}q \right\rangle \left[-\frac{\left\langle \bar{s}s \right\rangle \left\langle \bar{q}Gq \right\rangle m_s}{2^{8}\, \pi ^4}+\frac{\left\langle \bar{s}s \right\rangle \left\langle \bar{s}Gs \right\rangle m_s}{2^{9}\times 3 \, \pi ^4}+\left\langle GG \right\rangle^{2} \big(\frac{\left\langle \bar{s}Gs \right\rangle}{2^{14}\times 3 \, \pi ^6}-\frac{\left\langle \bar{q}Gq \right\rangle}{2^{13}\times 3   \, \pi ^6}\big)\right]\nonumber\\
&+&\frac{\left\langle \bar{q}Gq \right\rangle m_s \left(\left\langle \bar{s}s \right\rangle^2+6 \left\langle \bar{q}q \right\rangle^{2}\right)}{2^{9}\times 3 \, \pi ^4}-\frac{\left\langle \bar{q}Gq \right\rangle^2 m_s^2}{2^{13}\, \pi ^6}-\frac{\left\langle \bar{q}q \right\rangle^{2} \left(3 \left\langle \bar{s}Gs \right\rangle m_s+8 \pi ^2 \left\langle \bar{s}s \right\rangle^{2}\right)}{2^{8}\times 3^2 \, \pi ^4}\nonumber\\
&+&\frac{\left\langle \bar{q}q \right\rangle^3 \left\langle \bar{s}s \right\rangle}{2^{4}\times 3^2 \, \pi ^2}+\frac{\left\langle \bar{s}s \right\rangle \left\langle GG \right\rangle^{2} \left\langle \bar{q}Gq \right\rangle}{2^{14}\times 3 \, \pi ^6}\;,\\
c_0&=&\frac{\left\langle \bar{q}q \right\rangle^4 m_s^2}{2^{4}\times 3^2 \, \pi ^2}-\frac{\left\langle \bar{q}q \right\rangle^3 \left\langle \bar{s}s \right\rangle m_s^2}{2^{4}\times 3^2 \, \pi ^2}+\frac{\left\langle \bar{q}q \right\rangle^{2} \left\langle \bar{s}s \right\rangle^{2} m_s^2}{2^{6}\times 3^2 \, \pi ^2}\;,
\end{eqnarray}

\subsection{The spectral densities of $1^{--}$ $\Sigma$-$\bar{\Sigma}$ hexaquark states}
The $1^{--}$ $\Sigma$-$\bar{\Sigma}$ hexaquark state spectral densities on the OPE side can be expressed as:
\begin{eqnarray}
c_7&=&-\frac{1}{2^{20}\times 3^4\times 5^2 \times 7^2 \, \pi ^{10}}\;,\\
c_6&=&0\;,\\
c_5&=&\frac{\left\langle \bar{q}q \right\rangle m_s}{2^{15} \times 3 \times 5^2 \times 7 \, \pi ^8}-\frac{\left\langle \bar{s}s \right\rangle m_s}{2^{16} \times 3 \times 5^2 \times 7 \, \pi ^8}-\frac{\left\langle GG \right\rangle}{2^{21}\times 3 \times 5^2 \times 7 \, \pi ^{10}}\;,\\
c_4&=&-\frac{\left\langle \bar{q}Gq \right\rangle m_s}{2^{18}\times 3^2  \, \pi ^8}-\frac{\left\langle \bar{s}Gs \right\rangle m_s}{2^{17}\times 3 \times 5  \, \pi ^8}-\frac{\left\langle \bar{q}q \right\rangle^2}{2^{14}\times 3^2 \times 5\, \pi ^6}-\frac{\left\langle \bar{q}q \right\rangle \left\langle \bar{s}s \right\rangle}{2^{14}\times 3^3  \, \pi ^6}\;,\\
c_3&=&-\frac{\left\langle \bar{q}q \right\rangle^2 m_s^2}{2^{10}\times 3^2 \times 5 \, \pi ^6}+\left\langle \bar{q}q \right\rangle \bigg[\frac{\left\langle \bar{s}s \right\rangle m_s^2}{2^{10}\times 3^2 \times 5 \, \pi ^6}+\frac{\left\langle GG \right\rangle m_s}{2^{15} \times 3^2 \times 5 \, \pi ^8}+\frac{\left\langle \bar{q}Gq \right\rangle}{2^{13}\times 3^2 \, \pi ^6}\nonumber\\
&+&\frac{\left\langle \bar{s}Gs \right\rangle}{2^{11} \times 3^2 \times 5 \, \pi ^6}\bigg]-\frac{\left\langle \bar{s}s \right\rangle^2 m_s^2}{2^{12}\times 3^2 \times 5  \, \pi ^6}-\frac{\left\langle \bar{s}s \right\rangle \left\langle GG \right\rangle m_s}{2^{16}\times 3^2 \times 5  \, \pi ^8}+\frac{\left\langle \bar{s}s \right\rangle \left\langle \bar{q}Gq \right\rangle}{2^{11} \times 3^2 \times 5 \, \pi ^6}\;,\\
c_2&=&\frac{\left\langle \bar{q}q \right\rangle^3 m_s}{2^{7}\times 3^2  \, \pi ^4}+\left\langle \bar{q}q \right\rangle^2 \left[\frac{\left\langle \bar{s}s \right\rangle m_s}{2^{9}\times 3^2  \, \pi ^4}-\frac{\left\langle GG \right\rangle}{2^{13}\times 3^2 \, \pi ^6}\right]\nonumber\\
&+&\left\langle \bar{q}q \right\rangle \left[-\frac{\left\langle \bar{s}s \right\rangle^2 m_s}{2^{10}\times 3 \, \pi ^4}+\frac{\left\langle \bar{q}Gq \right\rangle m_s^2}{2^{13} \, \pi ^6}-\frac{\left\langle \bar{s}s \right\rangle \left\langle GG \right\rangle}{2^{15}\times 3  \, \pi ^6}\right]\nonumber\\
&+&\left\langle \bar{q}Gq \right\rangle \left[-\frac{\left\langle \bar{s}s \right\rangle m_s^2}{2^{14} \, \pi ^6}-\frac{\left\langle \bar{s}Gs \right\rangle}{2^{15}\, \pi ^6}\right]-\frac{\left\langle GG \right\rangle \left\langle \bar{q}Gq \right\rangle m_s}{2^{19} \, \pi ^8}-\frac{\left\langle \bar{q}Gq \right\rangle^2}{2^{14}\times 3  \, \pi ^6}\;,\\
c_1&=&\left\langle \bar{q}q \right\rangle^2 \left[-\frac{\left\langle \bar{q}Gq \right\rangle m_s}{2^{8} \, \pi ^4}-\frac{\left\langle \bar{s}Gs \right\rangle m_s}{2^{7}\times 3^2  \, \pi ^4}-\frac{\left\langle \bar{s}s \right\rangle^2}{2^{4}\times 3^3  \, \pi ^2}\right]+\frac{\left\langle \bar{s}s \right\rangle^2 \left\langle \bar{q}Gq \right\rangle m_s}{2^{8}\times 3^2 \, \pi ^4}\nonumber\\
&+&\left\langle \bar{q}q \right\rangle \left[-\frac{\left\langle \bar{s}s \right\rangle \left\langle \bar{q}Gq \right\rangle m_s}{2^{8}\times 3^2 \, \pi ^4}+\frac{\left\langle \bar{s}s \right\rangle \left\langle \bar{s}Gs \right\rangle m_s}{2^{8}\times 3^2 \, \pi ^4}+\left\langle GG \right\rangle \Big(\frac{\left\langle \bar{q}Gq \right\rangle}{2^{13}\times 3  \, \pi ^6}+\frac{\left\langle \bar{s}Gs \right\rangle}{2^{13}\times 3^2 \, \pi ^6}\Big)\right]\nonumber\\
&-&\frac{\left\langle \bar{q}Gq \right\rangle^2 m_s^2}{2^{12}\times 3 \, \pi ^6}-\frac{\left\langle \bar{q}q \right\rangle^3 \left\langle \bar{s}s \right\rangle}{2^{4}\times 3^2 \, \pi ^2}+\frac{\left\langle \bar{s}s \right\rangle \left\langle GG \right\rangle \left\langle \bar{q}Gq \right\rangle}{2^{13}\times 3^2 \, \pi ^6}\;,\\
c_0&=&\frac{\left\langle \bar{q}q \right\rangle^3 \left\langle \bar{s}s \right\rangle m_s^2}{2^{4}\times 3^2 \, \pi ^2}-\frac{\left\langle \bar{q}q \right\rangle^4 m_s^2}{2^{4}\times 3^2 \, \pi ^2}-\frac{\left\langle \bar{q}q \right\rangle^2 \left\langle \bar{s}s \right\rangle^2 m_s^2}{2^{6}\times 3^2 \, \pi ^2}\;,
\end{eqnarray}

\subsection{The spectral densities of $1^{++}$ $\Sigma$-$\bar{\Sigma}$ hexaquark states}
The $1^{++}$ $\Sigma$-$\bar{\Sigma}$ hexaquark state spectral densities on the OPE side can be expressed as:
\begin{eqnarray}
c_7&=&-\frac{1}{2^{20}\times 3^4\times 5^2 \times 7^2 \, \pi ^{10}}\;,\\
c_6&=&0\;,\\
c_5&=&\frac{\left\langle \bar{q}q \right\rangle m_s}{2^{15}\times 3 \times 5^2 \times 7  \, \pi ^8}-\frac{\left\langle \bar{s}s \right\rangle m_s}{2^{16}\times 3 \times 5^2 \times 7  \, \pi ^8}-\frac{\left\langle GG \right\rangle^{2}}{2^{21}\times 3 \times 5^2 \times 7 \, \pi ^{10}}\;,\\
c_4&=&-\frac{\left\langle \bar{q}Gq \right\rangle m_s}{2^{18}\times 3^2\, \pi ^8}-\frac{\left\langle \bar{q}q \right\rangle \left\langle \bar{s}s \right\rangle}{2^{14}\times 3^3  \, \pi ^6}+\frac{\left\langle \bar{q}q \right\rangle^{2}}{2^{14}\times 3^2 \times 5 \, \pi ^6}\;,\\
c_3&=&\left\langle \bar{q}q \right\rangle \left[\frac{2 \left\langle \bar{s}s \right\rangle m_s^2+\left\langle \bar{s}Gs \right\rangle}{2^{11}\times 3^2 \times 5   \, \pi ^6}+\frac{\left\langle GG \right\rangle^{2} m_s}{2^{15}\times 3^2 \times 5   \, \pi ^8}-\frac{\left\langle \bar{q}Gq \right\rangle}{2^{13}\times 3^2   \, \pi ^6}\right]-\frac{\left\langle \bar{s}s \right\rangle \left\langle GG \right\rangle^{2} m_s}{2^{16}\times 3^2 \times 5  \, \pi ^8}\nonumber\\
&-&\frac{m_s^2 (4 \left\langle \bar{q}q \right\rangle^{2}+\left\langle \bar{s}s \right\rangle^{2})}{2^{12}\times 3^2 \times 5  \, \pi ^6}+\frac{\left\langle \bar{s}s \right\rangle \left\langle \bar{q}Gq \right\rangle}{2^{11}\times 3^2 \times 5   \, \pi ^6}\;,\\
c_2&=&\left\langle \bar{q}q \right\rangle \left[\frac{\left\langle \bar{q}Gq \right\rangle m_s^2}{2^{13}\, \pi ^6}-\frac{m_s (8 \left\langle \bar{q}q \right\rangle^{2}+3 \left\langle \bar{s}s \right\rangle^{2})}{2^{10}\times 3^2   \, \pi ^4}-\frac{\left\langle \bar{s}s \right\rangle \left\langle GG \right\rangle^{2}}{2^{15}\times 3 \, \pi ^6}\right]-\frac{\left\langle \bar{q}Gq \right\rangle \left(2 \left\langle \bar{s}s \right\rangle m_s^2+\left\langle \bar{s}Gs \right\rangle\right)}{2^{15} \, \pi ^6}\nonumber\\
&+&\frac{5 \left\langle \bar{s}s \right\rangle \left\langle \bar{q}q \right\rangle^{2} m_s}{2^{9}\times 3^2  \, \pi ^4}-\frac{\left\langle GG \right\rangle^{2} \left\langle \bar{q}Gq \right\rangle m_s}{2^{19} \, \pi ^8}+\frac{\left\langle \bar{q}q \right\rangle^2 \left\langle GG \right\rangle^{2}}{2^{13}\times 3^2   \, \pi ^6}+\frac{\left\langle \bar{q}Gq \right\rangle^2}{2^{14}\times 3 \, \pi ^6}\;,\\
c_1&=&\left\langle \bar{q}q \right\rangle \left[-\frac{7 \left\langle \bar{s}s \right\rangle \left\langle \bar{q}Gq \right\rangle m_s}{2^{8}\times 3^2 \, \pi ^4}+\frac{\left\langle \bar{s}s \right\rangle \left\langle \bar{s}Gs \right\rangle m_s}{2^{8}\times 3^2 \, \pi ^4}+\left\langle GG \right\rangle^{2} \Big(\frac{\left\langle \bar{s}Gs \right\rangle}{2^{13}\times 3^2   \, \pi ^6}-\frac{\left\langle \bar{q}Gq \right\rangle}{2^{13}\times 3   \, \pi ^6}\Big)\right]\nonumber\\
&+&\frac{\left\langle \bar{q}Gq \right\rangle m_s \left(\left\langle \bar{s}s \right\rangle^2+9 \left\langle \bar{q}q \right\rangle^{2}\right)}{2^{8}\times 3^2 \, \pi ^4}+\frac{\left\langle \bar{q}Gq \right\rangle^2 m_s^2}{2^{12}\times 3 \, \pi ^6}-\frac{\left\langle \bar{s}Gs \right\rangle \left\langle \bar{q}q \right\rangle^{2} m_s}{2^{7}\times 3^2   \, \pi ^4}\nonumber\\
&+&\frac{\left\langle \bar{q}q \right\rangle^3 \left\langle \bar{s}s \right\rangle}{2^{4}\times 3^2 \, \pi ^2}-\frac{\left\langle \bar{q}q \right\rangle^2 \left\langle \bar{s}s \right\rangle^2}{2^{4}\times 3^3 \, \pi ^2}+\frac{\left\langle \bar{s}s \right\rangle \left\langle GG \right\rangle^{2} \left\langle \bar{q}Gq \right\rangle}{2^{13}\times 3^2   \, \pi ^6}\;,\\
c_0&=&\frac{\left\langle \bar{q}q \right\rangle^4 m_s^2}{2^{4}\times 3^2 \, \pi ^2}-\frac{\left\langle \bar{q}q \right\rangle^3 \left\langle \bar{s}s \right\rangle m_s^2}{2^{4}\times 3^2 \, \pi ^2}+\frac{\left\langle \bar{q}q \right\rangle^2 \left\langle \bar{s}s \right\rangle^2 m_s^2}{2^{6}\times 3^2 \, \pi ^2}\;,
\end{eqnarray}

\subsection{The spectral densities of $0^{-+}$ $\Xi$-$\bar{\Xi}$ hexaquark states}
The $0^{-+}$ $\Xi$-$\bar{\Xi}$ hexaquark state spectral densities on the OPE side can be expressed as:
\begin{eqnarray}
c_7&=&\frac{1}{2^{23}\times 3^2 \times 5^2 \times 7^2 \, \pi ^{10}}\;,\\
c_6&=&\frac{m_s^2   }{2^{20}\times 3^2 \times 5^2 \times 7 \, \pi ^{10}}\;,\\
c_5&=&\frac{\left\langle GG \right\rangle}{2^{22}\times 3^2 \times 5^2  \, \pi ^{10}}-\frac{\left\langle \bar{q}q \right\rangle m_s}{2^{16} \times 3^2 \times 5^2 \, \pi ^8}\;,\\
c_4&=&\left\langle \bar{q}q \right\rangle \left[\frac{\left\langle \bar{s}s \right\rangle}{2^{13}\times 3^2 \times 5  \, \pi ^6}-\frac{m_s^3}{2^{15}\times 3 \times 5  \, \pi ^8}\right]+\frac{\left\langle \bar{s}s \right\rangle m_s^3}{2^{16}\times 3 \times 5  \, \pi ^8}+\frac{\left\langle GG \right\rangle m_s^2}{2^{21}\times 3 \times 5  \, \pi ^{10}}\nonumber\\
&+&\frac{\left\langle \bar{q}Gq \right\rangle m_s}{2^{17}\times 3 \times 5  \, \pi ^8}+\frac{\left\langle \bar{s}Gs \right\rangle m_s}{2^{17}\times 3 \times 5  \, \pi ^8}+\frac{\left\langle \bar{s}s \right\rangle^2}{2^{14}\times 3^2 \times 5\, \pi ^6}\;,\\
c_3&=&\frac{\left\langle \bar{q}q \right\rangle^2 m_s^2}{2^{12}\times 3^2 \, \pi ^6}+\left\langle \bar{q}q \right\rangle \left[\frac{\left\langle \bar{s}s \right\rangle m_s^2}{2^{12}\times 3 \, \pi ^6}-\frac{\left\langle GG \right\rangle m_s}{2^{17}\times 3^2 \, \pi ^8}-\frac{\left\langle \bar{s}Gs \right\rangle}{2^{13}\times 3^2 \, \pi ^6}\right]-\frac{\left\langle \bar{s}s \right\rangle^2 m_s^2}{2^{13}\times 3^2 \, \pi ^6}\nonumber\\
&-&\frac{\left\langle \bar{s}s \right\rangle \left\langle GG \right\rangle m_s}{2^{18}\times 3^2  \, \pi ^8}+\left\langle \bar{q}Gq \right\rangle \left[\frac{m_s^3}{2^{15}\times 3  \, \pi ^8}-\frac{\left\langle \bar{s}s \right\rangle}{2^{13}\times 3^2 \, \pi ^6}\right]-\frac{\left\langle \bar{s}s \right\rangle \left\langle \bar{s}Gs \right\rangle}{2^{13}\times 3^2 \, \pi ^6}  \;,\\
c_2&=&\left\langle \bar{q}q \right\rangle \bigg[-\frac{\left\langle \bar{s}s \right\rangle^2 m_s}{2^{8}\times 3 \, \pi ^4}+\left\langle GG \right\rangle \Big(\frac{\left\langle \bar{s}s \right\rangle}{2^{13}\times 3^2 \, \pi ^6}-\frac{m_s^3}{2^{15}\times 3  \, \pi ^8}\Big)-\frac{\left\langle \bar{s}s \right\rangle m_s^4}{2^{10}\times 3 \, \pi ^6}\nonumber\\
&-&\frac{\left\langle \bar{q}Gq \right\rangle m_s^2}{2^{11}\times 3 \, \pi ^6}-\frac{\left\langle \bar{s}Gs \right\rangle m_s^2}{2^{11} \, \pi ^6}\bigg]+\left\langle \bar{q}q \right\rangle^2 \left[\frac{m_s^4}{2^{10}\times 3 \, \pi ^6}-\frac{\left\langle \bar{s}s \right\rangle m_s}{2^{7}\times 3^2  \, \pi ^4}\right]+\frac{\left\langle \bar{s}s \right\rangle^3 m_s}{2^{8}\times 3^2 \, \pi ^4}\nonumber\\
&+&\left\langle GG \right\rangle \left[\frac{\left\langle \bar{s}s \right\rangle m_s^3}{2^{16}\times 3 \, \pi ^8}+\frac{\left\langle \bar{q}Gq \right\rangle m_s}{2^{17}\times 3 \, \pi ^8}+\frac{\left\langle \bar{s}Gs \right\rangle m_s}{2^{16}\times 3 \, \pi ^8}+\frac{\left\langle \bar{s}s \right\rangle^2}{2^{13}\times 3^2 \, \pi ^6}\right]\nonumber\\
&+&\frac{\left\langle \bar{s}s \right\rangle^2 m_s^4}{2^{12}\times 3 \, \pi ^6}+\left\langle \bar{q}Gq \right\rangle \left[\frac{\left\langle \bar{s}Gs \right\rangle}{2^{13}\times 3  \, \pi ^6}-\frac{\left\langle \bar{s}s \right\rangle m_s^2}{2^{12} \, \pi ^6}\right]+\frac{\left\langle \bar{s}s \right\rangle \left\langle \bar{s}Gs \right\rangle m_s^2}{2^{11}\times 3 \, \pi ^6}+\frac{\left\langle \bar{s}Gs \right\rangle^2}{2^{14}\times 3  \, \pi ^6}\;,\\
c_1&=&\left\langle \bar{q}q \right\rangle^2 \left[-\frac{\left\langle \bar{s}s \right\rangle m_s^3}{2^{7}\, \pi ^4}+\frac{\left\langle \bar{s}Gs \right\rangle m_s}{2^{8}\times 3 \, \pi ^4}+\frac{\left\langle \bar{s}s \right\rangle^2}{2^{5}\times 3^2 \, \pi ^2}\right]-\frac{\left\langle \bar{s}s \right\rangle^2 \left\langle \bar{s}Gs \right\rangle m_s}{2^{8}\times 3 \, \pi ^4}+\frac{\left\langle \bar{q}Gq \right\rangle^2 m_s^2}{2^{13} \, \pi ^6}\nonumber\\
&+&\left\langle \bar{q}q \right\rangle \left[\frac{\left\langle \bar{s}s \right\rangle^2 m_s^3}{2^{8} \, \pi ^4}+\left\langle GG \right\rangle \Big(\frac{\left\langle \bar{s}s \right\rangle m_s^2}{2^{11}\times 3 \, \pi ^6}-\frac{\left\langle \bar{s}Gs \right\rangle}{2^{14}\times 3  \, \pi ^6}\Big)+\left\langle \bar{q}Gq \right\rangle \Big(\frac{\left\langle \bar{s}s \right\rangle m_s}{2^{7}\times 3 \, \pi ^4}-\frac{m_s^4}{2^{10}\, \pi ^6}\Big)\right] \nonumber\\
&+&\left\langle \bar{q}q \right\rangle \left[ \frac{3 \left\langle \bar{s}s \right\rangle \left\langle \bar{s}Gs \right\rangle m_s}{2^{9}\, \pi ^4}+\frac{\left\langle \bar{s}s \right\rangle^3}{2^{4}\times 3^2 \, \pi ^2}\right]\nonumber\\
&+&\left\langle GG \right\rangle \left[-\frac{\left\langle \bar{s}s \right\rangle^2 m_s^2}{2^{14}\times 3  \, \pi ^6}+\left\langle \bar{q}Gq \right\rangle \Big(\frac{m_s^3}{2^{16} \, \pi ^8}-\frac{\left\langle \bar{s}s \right\rangle}{2^{14}\times 3  \, \pi ^6}\Big)-\frac{\left\langle \bar{s}s \right\rangle \left\langle \bar{s}Gs \right\rangle}{2^{13}\times 3  \, \pi ^6}\right]\nonumber\\
&+&\left\langle \bar{q}Gq \right\rangle \left[\frac{\left\langle \bar{s}s \right\rangle^2 m_s}{2^{9}\, \pi ^4}+\frac{\left\langle \bar{s}s \right\rangle m_s^4}{2^{11} \, \pi ^6}+\frac{3 \left\langle \bar{s}Gs \right\rangle m_s^2}{2^{12} \, \pi ^6}\right]\;,\\
c_0&=&\left\langle \bar{q}q \right\rangle^2 \left[\frac{\left\langle \bar{s}s \right\rangle^2 m_s^2}{2^{2} \times 3^2 \, \pi ^2}+\frac{\left\langle \bar{s}Gs \right\rangle m_s^3}{2^{6}\times 3 \, \pi ^4}\right]+\frac{\left\langle \bar{s}s \right\rangle^4 m_s^2}{2^{6}\times 3^2 \, \pi ^2}-\frac{\left\langle \bar{s}s \right\rangle^2 \left\langle \bar{q}Gq \right\rangle m_s^3}{2^{9}\, \pi ^4}+\frac{\left\langle \bar{s}s \right\rangle^2 \left\langle \bar{s}Gs \right\rangle m_s^3}{2^{9}\times 3 \, \pi ^4}\nonumber\\
&+&\left\langle \bar{q}q \right\rangle \left[-\frac{\left\langle \bar{s}s \right\rangle^3 m_s^2}{2^{5}\times 3 \, \pi ^2}+\frac{\left\langle \bar{s}s \right\rangle \left\langle \bar{q}Gq \right\rangle m_s^3}{2^{7}\, \pi ^4}-\frac{\left\langle \bar{s}s \right\rangle \left\langle \bar{s}Gs \right\rangle m_s^3}{2^{8} \, \pi ^4}-\frac{\left\langle GG \right\rangle \left\langle \bar{s}Gs \right\rangle m_s^2}{2^{13} \, \pi ^6}\right]\nonumber\\
&+&\left\langle GG \right\rangle \left[\frac{\left\langle \bar{s}s \right\rangle \left\langle \bar{s}Gs \right\rangle m_s^2}{2^{13}\times 3  \, \pi ^6}-\frac{\left\langle \bar{s}s \right\rangle \left\langle \bar{q}Gq \right\rangle m_s^2}{2^{12}\times 3 \, \pi ^6}\right]+\frac{\left\langle \bar{q}Gq \right\rangle^2 m_s^4}{2^{12} \, \pi ^6}\;,
\end{eqnarray}

\subsection{The spectral densities of $0^{++}$ $\Xi$-$\bar{\Xi}$ hexaquark states}
The $0^{++}$ $\Xi$-$\bar{\Xi}$ hexaquark state spectral densities on the OPE side can be expressed as:
\begin{eqnarray}
c_7&=&-\frac{1}{2^{23}\times 3^2\times 5^2\times 7^2\, \pi ^{10}}\;,\\
c_6&=&\frac{  m_s^2}{2^{20}\times 3^2\times 5^2\times 7^1\, \pi ^{10}}\;,\\
c_5&=&\frac{\left\langle \bar{q}q \right\rangle m_s}{2^{16}\times 3^2 \times 5^2 \, \pi ^8}-\frac{\left\langle \bar{s}s \right\rangle m_s}{2^{15} \times 3^2 \times 5^2 \, \pi ^8}-\frac{\left\langle GG \right\rangle^{2}}{2^{22}\times 3^2\times 5^2\, \pi ^{10}}\;,\\
c_4&=&-\frac{\left\langle \bar{q}q \right\rangle \left(3 m_s^3+4 \pi ^2 \left\langle \bar{s}s \right\rangle\right)}{2^{15}\times 3^2 \times 5   \, \pi ^8}+\frac{6 \left\langle \bar{s}s \right\rangle m_s^3+3 \left\langle \bar{s}Gs \right\rangle m_s+8 \pi ^2 \left\langle \bar{s}s \right\rangle^{2}}{2^{17}\times 3^2 \times 5  \, \pi ^8}\nonumber\\
&+&\frac{\left\langle GG \right\rangle^{2} m_s^2}{2^{21}\times 3\times 5\, \pi ^{10}}-\frac{\left\langle \bar{q}Gq \right\rangle m_s}{2^{17}\times 3\times 5\, \pi ^8}\;,\\
c_3&=&\left\langle \bar{q}q \right\rangle \left[\frac{18 \left\langle \bar{s}s \right\rangle m_s^2+\left\langle \bar{s}Gs \right\rangle}{2^{13}\times 3^2   \, \pi ^6}+\frac{\left\langle GG \right\rangle^{2} m_s}{2^{17}\times 3^2  \, \pi ^8}\right]-\frac{7 \left\langle \bar{s}s \right\rangle \left\langle GG \right\rangle^{2} m_s}{2^{18}\times 3^2\, \pi ^8}\nonumber\\
&+&\frac{\left\langle \bar{q}Gq \right\rangle \left(3 m_s^3+4 \pi ^2 \left\langle \bar{s}s \right\rangle\right)}{2^{15}\times 3^2   \, \pi ^8}-\frac{m_s^2 (2 \left\langle \bar{q}q \right\rangle^{2}+\left\langle \bar{s}s \right\rangle^{2})+\left\langle \bar{s}s \right\rangle \left\langle \bar{s}Gs \right\rangle}{2^{13}\times 3^2   \, \pi ^6}\;,\\
c_2&=&\frac{64 \pi ^2 \left\langle \bar{s}s \right\rangle^3 m_s+3 \left(8 \left\langle \bar{s}s \right\rangle \left\langle \bar{s}Gs \right\rangle m_s^2+4 m_s^4 (4 \left\langle \bar{q}q \right\rangle^{2}+\left\langle \bar{s}s \right\rangle^{2})+\left\langle \bar{s}Gs \right\rangle^2\right)}{2^{14}\times 3^2 \, \pi ^6}\nonumber\\
&+&\left\langle \bar{q}q \right\rangle \bigg[-\frac{m_s \left(2 \left\langle \bar{s}s \right\rangle m_s^3+3 \left\langle \bar{s}Gs \right\rangle m_s+24 \pi ^2 \left\langle \bar{s}s \right\rangle^2\right)}{2^{11}\times 3  \, \pi ^6}-\frac{\left\langle GG \right\rangle^{2} \left(3 m_s^3+4 \pi ^2 \left\langle \bar{s}s \right\rangle\right)}{2^{15}\times 3^2   \, \pi ^8}\nonumber\\
&+&\frac{\left\langle \bar{q}Gq \right\rangle m_s^2}{2^{11}\times 3  \, \pi ^6}\bigg]+\left\langle GG \right\rangle^{2} \left[\frac{3 \left\langle \bar{s}s \right\rangle m_s^3+3 \left\langle \bar{s}Gs \right\rangle m_s+8 \pi ^2 \left\langle \bar{s}s \right\rangle^2}{2^{16}\times 3^2 \, \pi ^8}-\frac{\left\langle \bar{q}Gq \right\rangle m_s}{2^{17}\times 3 \, \pi ^8}\right]\nonumber\\
&-&\frac{\left\langle \bar{q}Gq \right\rangle \left(18 \left\langle \bar{s}s \right\rangle m_s^2+\left\langle \bar{s}Gs \right\rangle\right)}{2^{13}\times 3   \, \pi ^6}+\frac{\left\langle \bar{q}q \right\rangle^2 \left\langle \bar{s}s \right\rangle m_s}{2^{7}\times 3^2   \, \pi ^4}\;,\\
c_1&=&-\frac{\left\langle \bar{q}q \right\rangle^2 \left(30 \left\langle \bar{s}s \right\rangle m_s^3+3 \left\langle \bar{s}Gs \right\rangle m_s+8 \pi ^2 \left\langle \bar{s}s \right\rangle^2\right)}{2^{8}\times 3^2 \, \pi ^4}-\frac{\left\langle \bar{s}s \right\rangle^2 m_s \left(2 \left\langle \bar{s}s \right\rangle m_s^2+\left\langle \bar{s}Gs \right\rangle\right)}{2^{8}\times 3 \, \pi ^4}\nonumber\\
&+&\left\langle \bar{q}q \right\rangle \left[\frac{\left\langle \bar{s}s \right\rangle \left(54 \left\langle \bar{s}s \right\rangle m_s^3+45 \left\langle \bar{s}Gs \right\rangle m_s+32 \pi ^2 \left\langle \bar{s}s \right\rangle^2\right)}{2^{9}\times 3^2  \, \pi ^4}+\frac{\left\langle GG \right\rangle^{2} \left(16 \left\langle \bar{s}s \right\rangle m_s^2+\left\langle \bar{s}Gs \right\rangle\right)}{2^{14}\times 3 \, \pi ^6}\right] \nonumber\\
&-&\left\langle \bar{q}q \right\rangle \left[ \frac{\left\langle \bar{q}Gq \right\rangle \left(8 \pi ^2 \left\langle \bar{s}s \right\rangle m_s+3 m_s^4\right)}{2^{10}\times 3 \, \pi ^6}\right]+\frac{\left\langle \bar{q}Gq \right\rangle m_s \left(2 \left\langle \bar{s}s \right\rangle m_s^3+3 \left\langle \bar{s}Gs \right\rangle m_s+24 \pi ^2 \left\langle \bar{s}s \right\rangle^2\right)}{2^{12} \, \pi ^6}\nonumber\\
&+&\left\langle GG \right\rangle^{2} \left[\frac{\left\langle \bar{q}Gq \right\rangle \left(3 m_s^3+4 \pi ^2 \left\langle \bar{s}s \right\rangle\right)}{2^{16}\times 3 \, \pi ^8}-\frac{\left\langle \bar{s}s \right\rangle \left(7 \left\langle \bar{s}s \right\rangle m_s^2+2 \left\langle \bar{s}Gs \right\rangle\right)}{2^{14}\times 3 \, \pi ^6}\right]-\frac{\left\langle \bar{q}Gq \right\rangle^2 m_s^2}{2^{13}\, \pi ^6}\;,\\
c_0&=&\frac{\left\langle \bar{q}q \right\rangle^2 m_s^2 \left(3 \left\langle \bar{s}Gs \right\rangle m_s+32 \pi ^2 \left\langle \bar{s}s \right\rangle^2\right)}{2^{6}\times 3^2 \, \pi ^4}+\frac{\left\langle \bar{s}s \right\rangle^2 m_s^2 \left(3 \left\langle \bar{s}Gs \right\rangle m_s+8 \pi ^2 \left\langle \bar{s}s \right\rangle^2\right)}{2^{9}\times 3^2  \, \pi ^4}\nonumber\\
&+&\left\langle \bar{q}q \right\rangle \left[-\frac{\left\langle \bar{s}s \right\rangle m_s^2 \left(\left\langle \bar{s}Gs \right\rangle m_s+8 \pi ^2 \left\langle \bar{s}s \right\rangle^2\right)}{2^{8}\, \pi ^4}+\frac{5 \left\langle \bar{s}s \right\rangle \left\langle \bar{q}Gq \right\rangle m_s^3}{2^{7}\times 3 \, \pi ^4}-\frac{\left\langle GG \right\rangle^{2} \left\langle \bar{s}Gs \right\rangle m_s^2}{2^{13}\, \pi ^6}\right]\nonumber\\
&-&\frac{3 \left\langle \bar{s}s \right\rangle^2 \left\langle \bar{q}Gq \right\rangle m_s^3}{2^{9} \, \pi ^4}+\left\langle GG \right\rangle^{2} \left(\frac{\left\langle \bar{s}s \right\rangle \left\langle \bar{s}Gs \right\rangle m_s^2}{2^{13}\times 3   \, \pi ^6}-\frac{\left\langle \bar{s}s \right\rangle \left\langle \bar{q}Gq \right\rangle m_s^2}{2^{11}\times 3  \, \pi ^6}\right)+\frac{\left\langle \bar{q}Gq \right\rangle^2 m_s^4}{2^{12} \, \pi ^6}\;,
\end{eqnarray}
and
\begin{eqnarray}
\Pi^{sum}(M_B^2)=\frac{\left\langle \bar{q}q \right\rangle^2 \left\langle \bar{s}s \right\rangle^2 m_s^4}{2^{6}\times 3^2 \, \pi ^2}-\frac{\left\langle \bar{q}q \right\rangle \left\langle \bar{s}s \right\rangle^3 m_s^4}{2^{6}\times 3^2 \, \pi ^2}+\frac{\left\langle \bar{s}s \right\rangle^4 m_s^4}{2^{8}\times 3^2 \, \pi ^2}
\end{eqnarray}

\subsection{The spectral densities of $1^{--}$ $\Xi$-$\bar{\Xi}$ hexaquark states}
The $1^{--}$ $\Xi$-$\bar{\Xi}$ hexaquark state spectral densities on the OPE side can be expressed as:
\begin{eqnarray}
c_7&=&-\frac{1}{2^{20}\times 3^4\times 5^2 \times 7^2 \, \pi ^{10}}\;,\\
c_6&=&-\frac{ m_s^2 }{2^{20}\times 3^2 \times 5^2 \times 7 \, \pi ^{10}}\;,\\
c_5&=&\frac{\left\langle \bar{q}q \right\rangle m_s}{2^{15} \times 3 \times 5^2 \times 7 \, \pi ^8}+\frac{\left\langle \bar{s}s \right\rangle m_s}{2^{16}\times 3^2 \times 5^2 \times 7 \, \pi ^8}-\frac{\left\langle GG \right\rangle}{2^{21}\times 3 \times 5^2 \times 7 \, \pi ^{10}}\;,\\
c_4&=&\left\langle \bar{q}q \right\rangle \left[\frac{m_s^3}{2^{15}\times 3 \times 5  \, \pi ^8}-\frac{\left\langle \bar{s}s \right\rangle}{2^{14}\times 3^3  \, \pi ^6}\right] -\frac{\left\langle \bar{s}s \right\rangle m_s^3}{2^{16}\times 3 \times 5  \, \pi ^8}-\frac{\left\langle GG \right\rangle m_s^2}{2^{21}\times 3 \times 5  \, \pi ^{10}}\nonumber\\
&-&\frac{\left\langle \bar{q}Gq \right\rangle m_s}{2^{18}\times 3^2  \, \pi ^8}-\frac{\left\langle \bar{s}Gs \right\rangle m_s}{2^{17}\times 3 \times 5  \, \pi ^8}-\frac{\left\langle \bar{s}s \right\rangle^2}{2^{14}\times 3^2 \times 5\, \pi ^6}\;,\\
c_3&=&+\left\langle \bar{q}q \right\rangle \left[-\frac{\left\langle \bar{s}s \right\rangle m_s^2}{2^{11} \times 5\, \pi ^6}+\frac{\left\langle GG \right\rangle m_s}{2^{15} \times 3^2 \times 5 \, \pi ^8}+\frac{\left\langle \bar{s}Gs \right\rangle}{2^{11} \times 3^2 \times 5 \, \pi ^6}\right]\nonumber\\
&-&\frac{\left\langle \bar{q}q \right\rangle^2 m_s^2}{2^{10}\times 3^2 \times 5 \, \pi ^6}+\frac{\left\langle \bar{s}s \right\rangle^2 m_s^2}{2^{10}\times 3^2 \times 5 \, \pi ^6}+\frac{\left\langle \bar{s}s \right\rangle \left\langle \bar{s}Gs \right\rangle}{2^{13}\times 3^2 \, \pi ^6}\nonumber\\
&+&\frac{\left\langle \bar{s}s \right\rangle \left\langle GG \right\rangle m_s}{2^{15} \times 3^2 \times 5 \, \pi ^8}+\left\langle \bar{q}Gq \right\rangle \left[\frac{\left\langle \bar{s}s \right\rangle}{2^{11} \times 3^2 \times 5 \, \pi ^6}-\frac{m_s^3}{2^{15}\times 3  \, \pi ^8}\right]\;,\\
c_2&=&\left\langle \bar{q}q \right\rangle \bigg[\frac{5 \left\langle \bar{s}s \right\rangle^2 m_s}{2^{10}\times 3 \, \pi ^4}+\left\langle GG \right\rangle \left(\frac{m_s^3}{2^{15}\times 3  \, \pi ^8}-\frac{\left\langle \bar{s}s \right\rangle}{2^{15}\times 3  \, \pi ^6}\right)+\frac{\left\langle \bar{s}s \right\rangle m_s^4}{2^{10}\times 3 \, \pi ^6}+\frac{\left\langle \bar{q}Gq \right\rangle m_s^2}{2^{13} \, \pi ^6}\nonumber\\
&+&\frac{\left\langle \bar{s}Gs \right\rangle m_s^2}{2^{11} \, \pi ^6}\bigg]-\frac{\left\langle \bar{s}s \right\rangle^3 m_s}{2^{8}\times 3^2 \, \pi ^4}+\left\langle GG \right\rangle \left[-\frac{\left\langle \bar{s}s \right\rangle m_s^3}{2^{16}\times 3 \, \pi ^8}-\frac{\left\langle \bar{q}Gq \right\rangle m_s}{2^{19} \, \pi ^8}-\frac{\left\langle \bar{s}Gs \right\rangle m_s}{2^{16}\times 3 \, \pi ^8}-\frac{\left\langle \bar{s}s \right\rangle^2}{2^{13}\times 3^2 \, \pi ^6}\right]\nonumber\\
&+&\left\langle \bar{q}Gq \right\rangle \left(\frac{5 \left\langle \bar{s}s \right\rangle m_s^2}{2^{14} \, \pi ^6}-\frac{\left\langle \bar{s}Gs \right\rangle}{2^{15}\, \pi ^6}\right)-\frac{\left\langle \bar{s}s \right\rangle^2 m_s^4}{2^{12}\times 3 \, \pi ^6}\nonumber\\
&-&\frac{\left\langle \bar{s}s \right\rangle \left\langle \bar{s}Gs \right\rangle m_s^2}{2^{11}\times 3 \, \pi ^6}-\frac{\left\langle \bar{s}Gs \right\rangle^2}{2^{14}\times 3  \, \pi ^6}+\left\langle \bar{q}q \right\rangle^2 \left(\frac{\left\langle \bar{s}s \right\rangle m_s}{2^{9}\times 3 \, \pi ^4}-\frac{m_s^4}{2^{10}\times 3 \, \pi ^6}\right)\;,\\
c_1&=&\left\langle \bar{q}q \right\rangle^2 \left[\frac{5 \left\langle \bar{s}s \right\rangle m_s^3}{2^{6}\times 3^2 \, \pi ^4}-\frac{\left\langle \bar{s}Gs \right\rangle m_s}{2^{7}\times 3^2  \, \pi ^4}-\frac{\left\langle \bar{s}s \right\rangle^2}{2^{4}\times 3^3  \, \pi ^2}\right]+\frac{\left\langle \bar{s}s \right\rangle^3 m_s^3}{2^{8}\times 3^2 \, \pi ^4}\nonumber\\
&+&\left\langle \bar{q}q \right\rangle \left[-\frac{\left\langle \bar{s}s \right\rangle^2 m_s^3}{2^{6}\times 3 \, \pi ^4}+\left\langle GG \right\rangle \left(\frac{\left\langle \bar{s}Gs \right\rangle}{2^{13}\times 3^2 \, \pi ^6}-\frac{7 \left\langle \bar{s}s \right\rangle m_s^2}{2^{12}\times 3^2 \, \pi ^6}\right)\right] \nonumber\\
&+&\left\langle \bar{q}q \right\rangle\left\langle \bar{q}Gq \right\rangle \left[ \left(\frac{m_s^4}{2^{10}\, \pi ^6}-\frac{\left\langle \bar{s}s \right\rangle m_s}{2^{6}\times 3^2 \, \pi ^4}\right)-\frac{5 \left\langle \bar{s}s \right\rangle \left\langle \bar{s}Gs \right\rangle m_s}{2^{8}\times 3 \, \pi ^4}-\frac{\left\langle \bar{s}s \right\rangle^3}{2^{4}\times 3^2 \, \pi ^2}\right]\nonumber\\
&+&\left\langle GG \right\rangle \left[\frac{\left\langle \bar{s}s \right\rangle^2 m_s^2}{2^{13}\times 3  \, \pi ^6}+\left\langle \bar{q}Gq \right\rangle \left(\frac{\left\langle \bar{s}s \right\rangle}{2^{13}\times 3^2 \, \pi ^6}-\frac{m_s^3}{2^{16} \, \pi ^8}\right)+\frac{\left\langle \bar{s}s \right\rangle \left\langle \bar{s}Gs \right\rangle}{2^{13}\times 3  \, \pi ^6}\right]\nonumber\\
&+&\left\langle \bar{q}Gq \right\rangle \left[-\frac{\left\langle \bar{s}s \right\rangle^2 m_s}{2^{7}\times 3 \, \pi ^4}-\frac{\left\langle \bar{s}s \right\rangle m_s^4}{2^{11} \, \pi ^6}-\frac{3 \left\langle \bar{s}Gs \right\rangle m_s^2}{2^{12} \, \pi ^6}\right]+\frac{\left\langle \bar{s}s \right\rangle^2 \left\langle \bar{s}Gs \right\rangle m_s}{2^{8}\times 3 \, \pi ^4}-\frac{\left\langle \bar{q}Gq \right\rangle^2 m_s^2}{2^{12}\times 3 \, \pi ^6}\;,\\
c_0&=&\left\langle \bar{q}q \right\rangle^2 \left[-\frac{5 \left\langle \bar{s}s \right\rangle^2 m_s^2}{2^{4}\times 3^2 \, \pi ^2}-\frac{\left\langle \bar{s}Gs \right\rangle m_s^3}{2^{6}\times 3 \, \pi ^4}\right]\nonumber\\
&+&\left\langle \bar{q}q \right\rangle \left[\frac{\left\langle \bar{s}s \right\rangle^3 m_s^2}{2^{6}\, \pi ^2}-\frac{7 \left\langle \bar{s}s \right\rangle \left\langle \bar{q}Gq \right\rangle m_s^3}{2^{8}\times 3 \, \pi ^4}+\frac{\left\langle \bar{s}s \right\rangle \left\langle \bar{s}Gs \right\rangle m_s^3}{2^{8} \, \pi ^4}+\frac{\left\langle GG \right\rangle \left\langle \bar{s}Gs \right\rangle m_s^2}{2^{13} \, \pi ^6}\right]\nonumber\\
&-&\frac{\left\langle \bar{s}s \right\rangle^4 m_s^2}{2^{6}\times 3^2 \, \pi ^2}+\frac{3 \left\langle \bar{s}s \right\rangle^2 \left\langle \bar{q}Gq \right\rangle m_s^3}{2^{10}\, \pi ^4}-\frac{\left\langle \bar{s}s \right\rangle^2 \left\langle \bar{s}Gs \right\rangle m_s^3}{2^{9}\times 3 \, \pi ^4}\nonumber\\
&+&\left\langle GG \right\rangle \left[\frac{5 \left\langle \bar{s}s \right\rangle \left\langle \bar{q}Gq \right\rangle m_s^2}{2^{14}\times 3  \, \pi ^6}-\frac{\left\langle \bar{s}s \right\rangle \left\langle \bar{s}Gs \right\rangle m_s^2}{2^{13}\times 3  \, \pi ^6}\right]-\frac{\left\langle \bar{q}Gq \right\rangle^2 m_s^4}{2^{12} \, \pi ^6}\;,
\end{eqnarray}

\subsection{The spectral densities of $1^{++}$ $\Xi$-$\bar{\Xi}$ hexaquark states}
The $1^{++}$ $\Xi$-$\bar{\Xi}$ hexaquark state spectral densities on the OPE side can be expressed as:
\begin{eqnarray}
c_7&=&-\frac{1}{2^{20}\times 3^4\times 5^2 \times 7^2 \, \pi ^{10}}\;,\\
c_6&=&\frac{  m_s^2}{2^{20}\times 3^2\times 5^2\times 7^1\, \pi ^{10}}\;,\\
c_5&=&\frac{\left\langle \bar{q}q \right\rangle m_s}{2^{15}\times 3 \times 5^2 \times 7  \, \pi ^8}-\frac{13 \left\langle \bar{s}s \right\rangle m_s}{2^{16}\times 3^2\times 5^2 \times 7 \, \pi ^8}-\frac{\left\langle GG \right\rangle^{2}}{2^{21}\times 3 \times 5^2 \times 7 \, \pi ^{10}}\;,\\
c_4&=&-\frac{\left\langle \bar{q}q \right\rangle \left(9 m_s^3+10 \pi ^2 \left\langle \bar{s}s \right\rangle\right)}{2^{15} \times 3^3 \times 5 \, \pi ^8}+\frac{6 \left\langle \bar{s}s \right\rangle m_s^3+3 \left\langle \bar{s}Gs \right\rangle m_s+8 \pi ^2 \left\langle \bar{s}s \right\rangle^2}{2^{17}\times 3^2 \times 5  \, \pi ^8}\nonumber\\
&+&\frac{\left\langle GG \right\rangle^{2} m_s^2}{2^{21}\times 3\times 5\, \pi ^{10}}-\frac{\left\langle \bar{q}Gq \right\rangle m_s}{2^{18}\times 3^2\, \pi ^8}\;,\\
c_3&=&-\frac{\left\langle \bar{q}q \right\rangle^2 m_s^2}{2^{10}\times 3^2 \times 5  \, \pi ^6}+\left\langle \bar{q}q \right\rangle \left[\frac{21 \left\langle \bar{s}s \right\rangle m_s^2+\left\langle \bar{s}Gs \right\rangle}{2^{11}\times 3^2 \times 5   \, \pi ^6}+\frac{\left\langle GG \right\rangle^{2} m_s}{2^{15}\times 3^2 \times 5   \, \pi ^8}\right]-\frac{\left\langle \bar{s}s \right\rangle \left\langle GG \right\rangle^{2} m_s}{2^{13}\times 3^2 \times 5  \, \pi ^8}\nonumber\\
&+&\frac{\left\langle \bar{q}Gq \right\rangle \left(15 m_s^3+16 \pi ^2 \left\langle \bar{s}s \right\rangle\right)}{2^{15}\times 3^2 \times 5   \, \pi ^8}-\frac{\left\langle \bar{s}s \right\rangle \left(8 \left\langle \bar{s}s \right\rangle m_s^2+5 \left\langle \bar{s}Gs \right\rangle\right)}{2^{13}\times 3^2 \times 5  \, \pi ^6}\;,\\
c_2&=&\frac{\left\langle \bar{q}q \right\rangle^2 \left(2 \pi ^2 \left\langle \bar{s}s \right\rangle m_s+m_s^4\right)}{2^{10}\times 3 \, \pi ^6}+\left\langle GG \right\rangle^{2} \left[\frac{3 \left\langle \bar{s}s \right\rangle m_s^3+3 \left\langle \bar{s}Gs \right\rangle m_s+8 \pi ^2 \left\langle \bar{s}s \right\rangle^2}{2^{16}\times 3^2 \, \pi ^8}-\frac{\left\langle \bar{q}Gq \right\rangle m_s}{2^{19} \, \pi ^8}\right]\nonumber\\
&+&\left\langle \bar{q}q \right\rangle \bigg[-\frac{m_s \left(2 \left\langle \bar{s}s \right\rangle m_s^3+3 \left\langle \bar{s}Gs \right\rangle m_s+22 \pi ^2 \left\langle \bar{s}s \right\rangle^2\right)}{2^{11}\times 3  \, \pi ^6}-\frac{\left\langle GG \right\rangle^{2} \left(m_s^3+\pi ^2 \left\langle \bar{s}s \right\rangle\right)}{2^{15}\times 3 \, \pi ^8}\nonumber\\
&+&\frac{\left\langle \bar{q}Gq \right\rangle m_s^2}{2^{13}\, \pi ^6}\bigg]+\frac{64 \pi ^2 \left\langle \bar{s}s \right\rangle^3 m_s+3 \left(4 \left\langle \bar{s}s \right\rangle^2 m_s^4+8 \left\langle \bar{s}s \right\rangle \left\langle \bar{s}Gs \right\rangle m_s^2+\left\langle \bar{s}Gs \right\rangle^2\right)}{2^{14}\times 3^2 \, \pi ^6}\nonumber\\
&-&\frac{\left\langle \bar{q}Gq \right\rangle \left(22 \left\langle \bar{s}s \right\rangle m_s^2+\left\langle \bar{s}Gs \right\rangle\right)}{2^{15} \, \pi ^6}\;,\\
c_1&=&-\frac{\left\langle \bar{q}q \right\rangle^2 \left(42 \left\langle \bar{s}s \right\rangle m_s^3+3 \left\langle \bar{s}Gs \right\rangle m_s+8 \pi ^2 \left\langle \bar{s}s \right\rangle^2\right)}{2^{7} \times 3^3 \, \pi ^4}+\left\langle \bar{q}q \right\rangle \frac{\left\langle GG \right\rangle^{2} \left(22 \left\langle \bar{s}s \right\rangle m_s^2+\left\langle \bar{s}Gs \right\rangle\right)}{2^{13}\times 3^2   \, \pi ^6}\nonumber\\
&+&\left\langle \bar{q}q \right\rangle \left[\frac{\left\langle \bar{s}s \right\rangle \left(24 \left\langle \bar{s}s \right\rangle m_s^3+21 \left\langle \bar{s}Gs \right\rangle m_s+16 \pi ^2 \left\langle \bar{s}s \right\rangle^2\right)}{2^{8}\times 3^2 \, \pi ^4}-\frac{\left\langle \bar{q}Gq \right\rangle \left(16 \pi ^2 \left\langle \bar{s}s \right\rangle m_s+9 m_s^4\right)}{2^{10}\times 3^2   \, \pi ^6}\right]\nonumber\\
&+&\frac{\left\langle \bar{q}Gq \right\rangle m_s \left(6 \left\langle \bar{s}s \right\rangle m_s^3+9 \left\langle \bar{s}Gs \right\rangle m_s+64 \pi ^2 \left\langle \bar{s}s \right\rangle^2\right)}{2^{12}\times 3 \, \pi ^6}-\frac{\left\langle \bar{s}s \right\rangle^2 m_s \left(5 \left\langle \bar{s}s \right\rangle m_s^2+3 \left\langle \bar{s}Gs \right\rangle\right)}{2^{8}\times 3^2 \, \pi ^4}\nonumber\\
&+&\left\langle GG \right\rangle^{2} \left[\frac{\left\langle \bar{q}Gq \right\rangle \left(9 m_s^3+8 \pi ^2 \left\langle \bar{s}s \right\rangle\right)}{2^{16}\times 3^2 \, \pi ^8}-\frac{\left\langle \bar{s}s \right\rangle \left(3 \left\langle \bar{s}s \right\rangle m_s^2+\left\langle \bar{s}Gs \right\rangle\right)}{2^{13}\times 3   \, \pi ^6}\right]-\frac{\left\langle \bar{q}Gq \right\rangle^2 m_s^2}{2^{12}\times 3 \, \pi ^6}\;,\\
c_0&=&\frac{\left\langle \bar{q}q \right\rangle^2 m_s^2 \left(3 \left\langle \bar{s}Gs \right\rangle m_s+28 \pi ^2 \left\langle \bar{s}s \right\rangle^2\right)}{2^{6}\times 3^2 \, \pi ^4}-\frac{5 \left\langle \bar{s}s \right\rangle^2 \left\langle \bar{q}Gq \right\rangle m_s^3}{2^{10} \, \pi ^4}+\frac{\left\langle \bar{s}s \right\rangle^2 m_s^2 \left(3 \left\langle \bar{s}Gs \right\rangle m_s+8 \pi ^2 \left\langle \bar{s}s \right\rangle^2\right)}{2^{9}\times 3^2  \, \pi ^4}\nonumber\\
&+&\left\langle \bar{q}q \right\rangle \left[-\frac{\left\langle \bar{s}s \right\rangle m_s^2 \left(3 \left\langle \bar{s}Gs \right\rangle m_s+20 \pi ^2 \left\langle \bar{s}s \right\rangle^2\right)}{2^{8}\times 3 \, \pi ^4}+\frac{3 \left\langle \bar{s}s \right\rangle \left\langle \bar{q}Gq \right\rangle m_s^3}{2^{8}\, \pi ^4}-\frac{\left\langle GG \right\rangle^{2} \left\langle \bar{s}Gs \right\rangle m_s^2}{2^{13}\, \pi ^6}\right]\nonumber\\
&+&\left\langle GG \right\rangle^{2} \left[\frac{\left\langle \bar{s}s \right\rangle \left\langle \bar{s}Gs \right\rangle m_s^2}{2^{13}\times 3   \, \pi ^6}-\frac{7 \left\langle \bar{s}s \right\rangle \left\langle \bar{q}Gq \right\rangle m_s^2}{2^{14}\times 3 \, \pi ^6}\right]+\frac{\left\langle \bar{q}Gq \right\rangle^2 m_s^4}{2^{12} \, \pi ^6}\;,
\end{eqnarray}

\end{widetext}
\end{document}